\documentclass[10pt,journal,letterpaper,compsoc]{IEEEtran}

\usepackage{pgf}
\usepackage{amssymb,amsmath,bm}
\usepackage[lined,boxed,commentsnumbered,ruled]{algorithm2e}
\usepackage{graphicx}
\usepackage{stmaryrd}
\usepackage{multirow,url}
\usepackage{color,cite}
\usepackage{cite}
\usepackage{tikz,pgf}
\usetikzlibrary{fadings,shapes,arrows,shadows}
\usepackage{makecell}
\usepackage{array}
\usepackage{overpic} 
\usepackage{subfigure}
\usepackage{colortbl,textcomp,booktabs}

\newtheorem{lemma}{Lemma}

\newtheorem{proposition}{Proposition}
\newtheorem{definition}{Definition}
\newtheorem{theorem}{Theorem}

\definecolor{DGreen}{rgb}{0.16,0.38,0.27}
\definecolor{mygray}{gray}{.9}

\def\N{\mathcal{N}}	
\def\I{\mathcal{I}}

\def\Q{\mathcal{Q}}
\def\K{\mathcal{K}}

\def\cup{c_\textsc{up}}
\def\cdl{c_\textsc{dl}}

\def\etase{\mathcal{S}^\textsc{s}}
\def\etaex{\mathcal{S}^\textsc{r}}
\def\etanl{\mathcal{S}^\textsc{a}}
\def\ctx{s}
\def\wsp{p}
\def\delta{\Phi}

\ifCLASSINFOpdf
\else
\fi

\begin{document}
%
\title{Scalable Mobile Crowdsensing via Peer-to-Peer Data Sharing}
\author{Changkun~Jiang, 
        Lin~Gao,~\IEEEmembership{Senior~Member,~IEEE,}
        Lingjie~Duan,~\IEEEmembership{Member,~IEEE,}
        and~Jianwei~Huang,~\IEEEmembership{Fellow,~IEEE}
\IEEEcompsocitemizethanks{
\IEEEcompsocthanksitem  Part of the results appeared in IEEE GLOBECOM 2015 \cite{JGDH-GLOBECOM}.
\IEEEcompsocthanksitem Changkun Jiang and Jianwei Huang are with the Network Communications and Economics Lab, Department of Information Engineering, The Chinese University of Hong Kong, Shatin, N.T., Hong Kong, China. \protect\\
E-mail: \{jc012, jwhuang\}@ie.cuhk.edu.hk
\IEEEcompsocthanksitem Lin Gao is with the School of Electronic and Information Engineering, Harbin Institute of Technology (Shenzhen), Shenzhen, China. \protect\\
E-mail: gaolin@hitsz.edu.cn
\IEEEcompsocthanksitem Lingjie Duan is with the Engineering Systems and Design Pillar, Singapore University of Technology and Design, 8 Somapah Road, Singapore. \protect\\
E-mail: lingjie\_duan@sutd.edu.sg
}
}

\IEEEcompsoctitleabstractindextext{
\begin{abstract}
Mobile crowdsensing (MCS) is a new paradigm of sensing by taking advantage of the rich embedded sensors of mobile user devices.
However, the traditional server-client MCS architecture often suffers from the high operational cost on the centralized server (e.g., for storing and processing massive data), hence the poor scalability.
Peer-to-peer (P2P) data sharing can effectively reduce the server's cost by leveraging the user devices' computation and storage resources.
In this work, we propose a novel \emph{P2P-based MCS} architecture, where the sensing data is saved and processed in user devices locally and shared among users in a P2P manner.
To provide necessary incentives for users in such a system, we propose a {quality-aware} \emph{data sharing market}, where the users who sense data can \emph{sell} data to others who request data but not want to sense the data by themselves.
We analyze the user behavior dynamics from the game-theoretic perspective, and characterize the
existence and uniqueness of the game equilibrium.
We further propose best response iterative algorithms to reach the equilibrium with provable convergence.
Our simulations show that the P2P data sharing can greatly improve the social welfare, especially in the model with a high transmission cost and a low trading price.
\end{abstract}
}

\maketitle

\IEEEdisplaynontitleabstractindextext

\IEEEpeerreviewmaketitle


\IEEEraisesectionheading{\section{Introduction}\label{sec:introduction}}

\subsection{Background and Motivations}
\IEEEPARstart{M}{obile} Crowdsensing (MCS) has recently emerged as  a novel and fast-growing sensing paradigm, thanks to the proliferation of mobile devices (e.g., smartphones, tablets, and sensor-equipped vehicles) and their embedded diverse mobile sensors. In MCS, the target sensing data are collected by a large group of mobile users using their mobile devices.
Due to the low deploying cost and the high sensing coverage, MCS has been implemented in a broad range of applications such as urban dynamic mining, public safety, and environment monitoring \cite{waze,opensignal,UrbanAtmosphere}.

The existing MCS applications mainly rely on the centralized \emph{server-client} architecture \cite{SmartphoneCollaboration,TieLuo,XZhang,LinGao}, where the participating users (clients) sense and report the data to a central  server, who further processes and distributes the data to those users who request to access the data.
However, the centralized architecture may not be suitable for those application scenarios with a large number of users and/or a large number of data requests, due to the high operational cost on the server (e.g., for data exchanging, processing, and storage).
For example, Niwa \emph{et al.}~in  \cite{MPSDataStore} have illustrated an example based on real testbed measurements, where 25 million smartphones sense data simultaneously, each collecting 1 Byte of data every minute and uploading to a storage server. In that case, the server needs to have 3Gbps of communication bandwidth and  enough storage to accommodate 1,350GB  per hour (hence 12PB per year), and needs to manage nearly 36 billion sensor data reports in one hour.
Meanwhile, mobile devices are becoming increasingly powerful, and their computation and storage capabilities are often under-utilized.
Furthermore, the emerging peer-to-peer sharing economy paradigms, in particular, Clone2Clone for smartphone connection \cite{C2C} and P2P Cloud \cite{P2PCloud}, enable the efficient peer-to-peer (P2P) data sharing by leveraging the sensing and processing capabilities of mobile devices.
This motivates us to shift part of the computation and storage burden on the server to the distributed mobile devices, giving rise to a more scalable architecture of mobile crowdsensing with P2P data sharing \cite{CrowdsourcingSmartphones,SmartphoneSocialNetworks, locationP2P}.

Specifically, in a P2P-based MCS system, the sensing data may not be reported to and saved in the server; instead, they can be saved and processed in mobile users' devices distributedly (via some mobile apps or dedicated middlewares as in \cite{CrowdsourcingSmartphones}) and shared among users directly.
The functionality of the server, similar as in traditional P2P networks, is mainly to keep track of each user's data occupancy information (e.g., which data she has) and network connection information (e.g., IP address of each user). With such information, the server can help users connect and share data with each other.
Moreover, data sharing among users can be done based on the local interactions (e.g., via WiFi or Bluetooth) when they are close enough, or directly through the Internet when they are not locally connected.

\begin{figure}
\centering
\definecolor{lavander}{cmyk}{0,0.48,0,0}
\definecolor{violet}{cmyk}{0.79,0.88,0,0}
\definecolor{burntorange}{cmyk}{0,0.52,1,0}

\def\lav{green!50!black}
\def\oran{red!30}

\tikzstyle{peers}=[draw,circle,\lav,bottom color=\lav,
                  top color= white, text=violet,minimum width=10pt]
\tikzstyle{superpeers}=[draw,circle,blue, left color= blue,
                       text=blue,minimum width=10pt]
\tikzstyle{aliens}=[draw,circle,gray,bottom color=gray,
 top color= white, text=gray,minimum width=10pt]
\usetikzlibrary{shapes,shadows,calc}
\usepgflibrary{arrows}
\begin{tikzpicture}[auto, thick]
\def\aboxl[#1,#2,#3,#4,#5]#6{%
  \node[draw, cylinder, alias=cyl, shape border rotate=90, aspect=#3, %
  minimum height=#1, minimum width=#2, outer sep=-0.5\pgflinewidth, %
  color=orange!40!black, left color=orange!70, right color=orange!80, middle color=white] (#4) at #5 {};%
  \node at #5 {#6};%
  \fill [orange!30] let \p1 = ($(cyl.before top)!0.5!(cyl.after top)$), \p2 =
  (cyl.top), \p3 = (cyl.before top), \n1={veclen(\x3-\x1,\y3-\y1)},
  \n2={veclen(\x2-\x1,\y2-\y1)} in (\p1) ellipse (\n1 and \n2); }

\aboxl[45,40,1.6,database,(-5,0)] {$\mathcal{S}$};

  \node[cloud, fill=gray!20, cloud puffs=16, cloud puff arc= 100,
        minimum width=5cm, minimum height=3cm, aspect=1] at (0,0) {};

  \foreach \place/\name in {{(0,-0.3)/a}, {(1,0)/b}, {(1,1)/c}, {(0,0.3)/d},{(-1,0)/e}}
    \node[superpeers] (\name) at \place {};
   %
  \foreach \pos/\i in {above left of/1, left of/2, below left of/3}
    \node[peers, \pos = e] (e\i) {};
   \foreach \speer/\peer in {e/e1,e/e2,e/e3}
    \path [->](\speer) edge (\peer);
   \foreach \pos/\i in {above right of/1, right of/2, below right of/3}
    \node[peers, \pos =b ] (b\i) {};
   \foreach \speer/\peer in {b/b1,b/b2,b/b3}
   \path [->](\speer) edge (\peer);
   \node[peers, above of=d] (d1){};
   \path[->] (d) edge (d1);
   \foreach \pos/\i in {below left of/1, below of/2}
   \node[peers, \pos =a ] (a\i) {};
   \foreach \speer/\peer in {a/a1,a/a2}
   \path [->](\speer) edge (\peer);
\node[aliens] () at (-0.5,0.8) {};
\node[aliens] () at (0.5,-0.8) {};
\node[aliens] () at (1,-0.8) {};
\draw [<->,thick,dashed] (database) -- (-2.5,0);
\node () at (-3.5,0.3) {{\scriptsize Query}};
\node () at (-5,-1.7) {{\scriptsize Server}};
\node () at (0,-1.7) {{\scriptsize P2P-based Crowdsensing Network}};
\end{tikzpicture}
\caption{P2P-based Mobile Crowdsensing Model.
Blue users: sensing data and sharing data with other users;
Green users: requesting data from other users who sense data;
Gray users: doing nothing.
}
\label{figure_P2PMCS}
\end{figure}
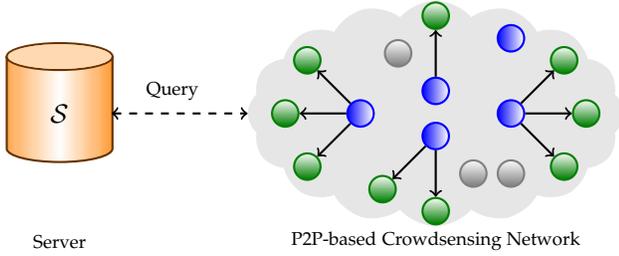

Fig. \ref{figure_P2PMCS} illustrates such a P2P-based MCS model, where the blue users sense data and share the sensing data with green users (via local WiFi or the Internet),
and the server is only responsible for the necessary control signal   exchange with users (e.g., establishing a network connection between  users).
There are   several commercial or demonstrative P2P-based MCS systems in practice, including MPSDataStore \cite{MPSDataStore}, SmartP2P \cite{CrowdsourcingSmartphones,SmartphoneSocialNetworks}, and LL-Net \cite{locationP2P}. The main technical challenges of deploying such a system include  tracking the data stored in mobile devices efficiently and conducting the on-demand requests scheduling and sensor data transmission effectively.

Several recent works have been devoted to studying the scalable system design for P2P-based MCS, by using the hierarchical structure and distributed data sharing among mobile devices \cite{App0,App1,App2,App3,App4}.
The key idea is to offload some or all of the computation and storage requirements to mobile devices.
However, these studies focused only on the technical issues in P2P-based MCS such as how to store data distributedly and how to search the distributed data efficiently, and none of them considered the \emph{economics issues} in such a system.
Due to the cost involved in data sensing and sharing, users may not have incentives to collaborate with each other without proper economic incentives.
This motivates our study of the \emph{economic incentive} issue in the P2P-based MCS system here.

We would like to emphasize that the incentive mechanisms for traditional P2P content distribution systems (e.g., \cite{P2Psharing,Buragohain}) are \emph{not} directly applicable for the  P2P-based MCS system.
First, traditional P2P systems usually assumed that users (peers) are endowed with different contents exogenously, and focused on the distribution of contents among users.
In our P2P-based MCS system, however, a distinctive feature is the joint consideration of the endogenous data generation (sensing) and distribution (sharing). This will significantly complicate the incentive design, as users have more than one way to obtain data, i.e., sensing themselves and requesting from others.
Second, traditional P2P systems usually assumed that each content is associated with a fixed quality.
In our P2P-based MCS system, however, a data can be selectively collected with different \emph{qualities}.
In reality, the population of mobile devices, the type of data each can produce, and the quality in terms of accuracy depend on many factors such as the user mobility, the communication channel variation, the device state (e.g., energy level), and the user preference \cite{survey}.
This heterogeneity in terms of mobile devices and their users' preferences leads to the tradeoff between data quality and resource consumptions. For example, location data can be obtained through using GPS, WiFi, and cellular networks, with decreasing levels of accuracy. Compared to WiFi and cellular networks, continuous GPS location sampling provides the most accurate location information, while draining the battery faster.
Hence, the consideration of data quality and the associated cost is very important for designing good incentive mechanisms to lead to good overall system performance \cite{Quality1,Quality2,Quality3}. The key difference is that we consider the novel peer-to-peer data sharing scenario to save the crowdsensing server congestion cost, while the above works \cite{Quality1,Quality2,Quality3} require the server-customer service interaction without data sharing among users on the network edge. Such a model difference leads to very different modeling, analysis, and solution.

\subsection{Solutions and Contributions}

In this work, we focus on the incentive design and economic analysis for the P2P-based MCS system with the quality-aware data sharing.
To achieve this goal, we propose a {quality-aware data sharing market},
where each user can choose to be a \emph{data
sensor}, sensing data with a specified quality and sharing the data with others (with certain reward),
or a \emph{data requester}, requesting data of the desired quality from a data sensor (with certain payment) instead of sensing data by herself.
Obviously, data sensors are \emph{sellers} and data requesters are \emph{buyers} in the data sharing market.

We aim to answer the following two important questions in such a data sharing market:
\begin{itemize}
\item How to design   proper market mechanisms to incentivize users to participate in the P2P-based MCS system and choose the desirable behaviors?
\item What is the equilibrium point of such a data market?
\end{itemize}
The former one is related to the mechanism design, i.e., designing the rules of the market.
The latter one is related to the game theoretical analysis, i.e., studying
how the market would evolve under the proposed rules.
First, we propose a  {general pricing scheme} for the data sharing (trading) among data sensors (sellers) and data requesters (buyers), which combines both the  {revenue sharing scheme} and the \emph{quality-based} pricing scheme.
With the proposed pricing scheme, the reward for a data sensor from selling data to a data requester   consists of (i) a portion of the total benefit that the requester achieves (from consuming the data), and (ii) a quality-aware data price.

Then, we perform the {game-theoretic analysis} for the data sharing market under the proposed pricing scheme.
In particular, we analyze the user behaviors and strategic interactions under the above pricing scheme systematically, for both scenarios of the quality-unaware data sharing and the quality-aware data sharing (capturing whether the data can be sensed and sold in different qualities).
We characterize the conditions for the market equilibrium, and prove the existence and
uniqueness of the equilibrium.
We further propose a generalized best response iterative algorithm that guarantees to converge to the market equilibrium.

Finally, we further model and analyze the \emph{cross-quality} data sharing, where a high quality
data can be transformed into a low-quality one and shared with a low-quality data requester.
This is quite often in practice, as most data (e.g., photo or video) can be easily transformed to a lower quality one through down sampling.
We analyze how the cross-quality sharing in P2P-based MCS affects the user behaviors as well as the market equilibrium.

The main results and key contributions  of this paper are summarized as follows.
\begin{itemize}
\item \emph{Novel P2P-based MCS Model:}
To our best knowledge, this is the first work that comprehensively analyzes the economic incentive issue in an MCS system with P2P-based data sharing, which is important for addressing the increasingly important scalability issue in MCS systems.

\item \emph{Market Mechanism Design and Game-theoretic Analysis:}
We propose a quality-aware data sharing market, together with a general data pricing scheme,
and
analyze the user behavior and market equilibrium systematically from the game-theoretic perspective.
Such an equilibrium analysis can help us to understand how the market evolves and what is the likely market outcome.

\item \emph{Observations and Insights:}
Our results show that the ratio of the equilibrium social welfare to the maximum social welfare benchmark increases with the data transmission cost and decreases with the data trading price, which implies that
the P2P-based MCS model is most effective when the transmission cost is high and the trading price is low.
We further show that the cross-quality sharing will drive more data sensors to sense the higher quality data, yet it does not have a significant impact on the data requesters'
quality selections or the achieved equilibrium social welfare, comparing with the scenario where cross-quality data sharing is not allowed.
\end{itemize}

The rest of the paper is organized as follows.
In Section~\ref{sec_model}, we present the system model. In Section~\ref{sec_MarketEvolutionAnalysis}, we analyze the quality-unaware game equilibrium.
In Section \ref{sec:Quality}, we analyze the quality-aware game equilibrium.
We present the simulation results in Section \ref{sec:NumericalResults} and conclude in Section~\ref{sec:conclusion}.


\section{System Model}\label{sec_model}
In this section, we first introduce the network, data, and user models. Then, we provide the pricing scheme and model the corresponding users' payoffs.
Finally, we formulate the users' strategic interactions as a non-cooperative game.

\subsection{Network Model}

We consider a quality-aware P2P-based MCS model with a set $\N=\{1, \cdots,N\}$ of mobile users, who can sense some data in a certain area and share the sensing data with each other in a distributed P2P manner.
Each data refers to a piece of specific information at a particular location and time.\footnote{For example, the data can be the cellular/WiFi signal strength of a particular region in OpenSignal \cite{opensignal}.  Users can obtain such data by
sampling the cellular/WiFi interfaces of their mobile devices. However, constantly sampling will incur a high battery consumption
of the mobile device. If another user has already sensed the signal strength with a more energy-efficient device, a
rational user would prefer to request the data from the energy-efficient device rather than sense by herself.}
Each user has the potential to sense a specific area consisting of one or multiple data,
depending on factors such as her mobility, device type, and energy budget.
We consider a set $\I=\{1, \cdots,I\}$ of different data, which can be used by one or multiple sensing applications. We consider multiple data markets corresponding to multiple locations, each of which runs independently at the same time. Hence, we focus on the operation of a single data market at a single location.
The data $i \in \I $ is associated with a weight~$w_{i}$, capturing the importance of the data.
For example, a hotspot data often has a larger weight than a non-hotspot data.\footnote{Here, hotspot means that some data at a particular location and time are more critical to users, e.g., the crowdedness of a shopping mall at weekend or a key road segment on a weekday.}

Each user can obtain her interested data in two ways: (i) acting as a data \emph{sensor} and sensing data directly, or (ii) acting as a data \emph{requester} and requesting data from a data sensor.
The latter case may happen when the user is not able to sense the data by herself (e.g., due to the mobility or device capability constraint), or when the user's sensing cost is very large (e.g., due to the energy budget constraint).
The data sharing among a data sensor and requester can be based on local WiFi or Bluetooth connections or the Internet connection.\footnote{Our model shares some similarity with the CrowdWatch model in \cite{App0}, with the key idea of leveraging on-demand data sharing among mobile devices to alleviate the burden of the server. The key difference is that \cite{App0} is a system paper without theoretical analysis, and only considers local data sharing using short-range communications. Hence, the encounter probability of users needs to be modeled if we aim to analyze the system; Unlike\cite{App0}, our theoretical analysis also allows the remote data sharing through the Internet for common data interests, and this feature is characterized by the transmission cost in our model.}
To facilitate such data sharing, the server needs to keep track of each user's network connection information (e.g., IP address) and data occupancy information (e.g., which data she has)  similar as in the traditional P2P system \cite{P2Psharing,Buragohain}.
 Note that the server does not need to store and process the sensing data.
{All the data processing and aggregations will be conducted by the associated apps on the mobile devices.}

\subsection{Data Quality}
A data can be captured by different \emph{qualities} (e.g., a photo can be captured by different resolutions).
Similar as in \cite{Quality2}, we consider that each data has $K$ types of discrete quality grades, indexed by $\mathcal{K}=\{1, \cdots,K\}$. Let $q_k\in\mathcal{Q} \triangleq \{q_k:k\in\mathcal{K}\}$ denote the $k$-th quality for the data.
Without loss of generality, we assume that $0  < q_1 < \cdots < q_K $.
The quality of a user's data can be effectively inferred by using similar methods in \cite{Quality1,Quality2,Quality3}.

In general, a user needs to consume more resources (hence incurs a higher sensing cost) for sensing a data with a higher quality.
Due to the user heterogeneity, different users may incur different sensing costs for sensing the same data with the same quality.
Moreover, a user often prefers a data with a higher quality than with a lower quality.
Similarly, different users may have different personal preferences for the same data with the same quality.

\subsection{User Model}\label{subsec_PricingSchemes}
When user $n\in\N$ obtains the data $i \in \I$ of quality $q_k$  (through either sensing or purchasing the data), she obtains a utility of $u_{ni}(q_k)$. In this work we adopt the following linear value function:
 \begin{equation}\label{eq:value}
u_{ni}(q_k)=v_{ni} \cdot f(q_k),
 \end{equation}
where
$v_{ni}$ is a factor evaluating the user's preference for different qualities,
and $f(\cdot)$ is an increasing function of $q_k$.

A user incurs a cost when sensing data.
Let $b_{ni}(q_k)$ denote user $n$'s  sensing cost for sensing the data $i\in\I$ with quality $q_k$.
In this work, we adopt the following linear sensing cost function:
 \begin{equation}\label{eq:cost}
b_{ni}(q_k)=c_{ni} \cdot g(q_k),
 \end{equation}
where
$c_{ni}$ is a factor evaluating the user's sensing cost for different qualities,
and $g(\cdot)$ is an increasing function of $q_k$.

Furthermore, when sharing data between two users (i.e., a sensor and a requester), there will be some \emph{data transmission cost}, mainly including the data upload cost (to the Internet) for the sensor and the data downloading cost (from the Internet)  for the requester.
For convenience, we assume that on average, all users have the same data uploading cost $\cup $ and downloading cost $\cdl $ for any data with any cost.
Hence, the total transmission cost for  the sharing of one data between two users is
 \begin{equation}
\ctx=\cdl  + \cup .
 \end{equation}
This is reasonable when different data with different qualities have approximately the same size, hence leads to the same transmission cost when sharing between users. In the future we will consider the case where quality has a significant impact on the data size and hence the data transmission cost.

Based on the above, we can see that each user $n $ can be fully characterized by the sensing cost factors $\{ c_{ni}, \forall i\in\I \}$ and the value factors $\{v_{ni}, \forall i\in\I \}$, as all other parameters (e.g., $ s$, $ f(\cdot)$, and  $g(\cdot)$) are identical for all users.
Without loss of generality, in the following analysis, \textbf{we focus on the operation for a particular data $i \in \I$}.\footnote{In this work, we do not consider the correlation across different data. This is often true for the applications with a low data correlation (e.g., taking photos of different buildings at different locations). In a more general case where the application requires correlated data across different locations and times (e.g., the air quality of a city during a day), we need to consider the sharing of different data jointly. We will leave this analysis in our future work.}
Hence, each user   $n $ can be fully characterized by a sensing cost evaluation factor $ c_{ni}$ and a value evaluation factor $v_{ni}$.

\subsection{Pricing Scheme}\label{sec:model:pricing}
When a data sensor shares the sensing data with a requester, the requester needs to pay the sensor.
Such a compensation can be related to the benefit that the requester achieves (from consuming the data) or the cost that the sensor incurs (for sensing the data).
The former one corresponds to the \emph{revenue sharing scheme},
and the latter one corresponds to the \emph{quality-based pricing scheme}, both widely used in reality \cite{pricingschemes}.

In this work, we adopt a general pricing scheme, which combines both the revenue sharing scheme and the quality-based pricing scheme. Formally,
\begin{definition}[General Pricing Scheme]
Suppose that a data requester achieves a total benefit $z$ from the data with quality $q_k$.
Then, the requester will pay the corresponding data sensor
\begin{equation}\label{eq:price}
z \cdot (1 - \eta) + \wsp \cdot h(q_k),
\end{equation}
where $\eta \in [0,1]$ is the revenue sharing factor, $h(q_k)$ is an increasing function of $q_k$, and $\wsp \cdot h(q_k)$ is the quality-aware price.
\end{definition}

It is easy to see that the above pricing scheme includes both the pure revenue sharing scheme (with $\wsp = 0$) and the pure quality-based pricing scheme  (with $\eta=1$) as special cases.
The key notations in this paper are listed in Table~\ref{tab:notations}.
\begin{table}[!t]
\setlength{\tabcolsep}{1pt}
\renewcommand{\arraystretch}{1.0}
\caption{Key Notations}\label{tab:notations}
\centering
\begin{tabular}{>{\scriptsize}c>{\scriptsize}c}
\toprule
{\bf Symbols} & {\bf Physical Meaning}\\
\midrule
$\N=\{1,\cdots,N\}$ & Set of mobile users\\
\rowcolor{mygray}
$\I=\{1,\cdots,I\}$ & Set of data\\
$\K=\{1,\cdots,K\}$ & Set of data quality types\\
\rowcolor{mygray}
$\Q=\{q_k: k\in\K\}$ & Set of data qualities\\
$u_{ni}(q_k)$ & Value of user $n$ with quality $q_k$ for data $i$\\
\rowcolor{mygray}
$b_{ni}(q_k)$ & Sensing cost of user $n$ with $q_k$  for data $i$\\
$v_{ni}$ or $v$ & Marginal user value with respect to data quality\\
\rowcolor{mygray}
$c_{ni}$ or $c$ & Marginal sensing cost with respect to data quality\\
$s$ & Total transmission cost of sharing one data item\\
\rowcolor{mygray}
$\eta$ & Revenue sharing factor\\
$\wsp \cdot h(q_k)$ & Quality-based pricing\\
\rowcolor{mygray}
$x\in\{\textsc{s},\textsc{r},\textsc{a}\}$ & Users' roles as sensor, requester, and alien\\
$\pi_{vc}(x,q_k)$  & User payoff choosing role $x $ and quality $q_k$\\
\rowcolor{mygray}
$\Phi_k$ & Average sharing benefit of a sensor choosing quality $q_k$\\
$\etase_k$, $\etaex_k$, and $\etanl $ & Sets of sensors with $q_k$, requesters with $q_k$, and aliens\\
\rowcolor{mygray}
$B^{\textsc{se}}_k$  & Total sharing benefit provided by all requesters with $q_k$\\
$ N^{\textsc{se}}_k$ & Total number of users choosing to be sensors with $q_k$\\
\rowcolor{mygray}
$\Lambda_k (\boldsymbol{\Phi} )$ & Functions of $\Phi_k$ to compute the equilibrium of $\Phi_k$ \\
\bottomrule
\end{tabular}
\end{table}

\subsection{User Behavior}\label{subsec:payoff}

To obtain the data (a data $i\in\I$), a user can choose to be a \emph{data sensor} (who senses the data directly) or a \emph{data requester} (who requests the data from a sensor).
The user can also choose to be an \emph{alien} (who neither senses nor requests data).
More specifically,
\begin{itemize}
\item \emph{Sensor}:
As a data sensor, the user senses the data directly with a specified quality $q_k\in\Q$  with some  sensing cost (e.g., the energy cost).
Meanwhile, the user can potentially share (sell) the sensing data with others to obtain some reward;

\item \emph{Requester}:
As a data requester, the user requests data with a desirable quality $q_k\in\Q$ from a sensor who has sensed the data already. Such sharing  introduces some data transmission cost.
The requester needs to bear all of the transmission cost, and provide some additional reward to the sensor.

\item \emph{Alien}:
As an alien, the user neither senses the data nor requests data from others.
This may occur when the user is not interested in the data or the cost of obtaining the data is too high.
\end{itemize}
It is notable that a user can choose different roles for different data, e.g.,  be a data sensor for one data while a data requester or alien for another data.

Without loss of generality, we consider a generic user $n \in \N$ (for data $i$).
For presentation convenience, \emph{we omit the subscripts
$n$ and $i$ whenever there is no confusion}, hence we can write the parameters $c_{ni}$ and  $v_{ni}$ as $c$ and $v$, respectively.
As mentioned previously, users are fully characterized by $c$ and $v$, and
different users may have different $c$ and $v$.
For convenience, we will use $(v,c)$ to characterize the user type.
For simplicity, we assume that both $v$ and $c$ follow independent uniform distributions over $[0,1]$ across all users, and denote the joint distribution by $\zeta_{vc}(v,c)$.\footnote{The assumption of the uniform distribution is mainly used for deriving analytical solutions and obtaining clear insights. Our analysis procedure is still applicable under more general distributions of $v$ and $c$, but we may not obtain closed-form solutions. Sometimes it is possible to prove the properties under general distribution using implicit function theorem (such as in our earlier work\cite{Lingjie}), but more often we need to rely on numerical methods to understand the existence and uniqueness of the game equilibrium.}

Let $x\in\{\textsc{s},\textsc{r},\textsc{a}\}$ denote the \emph{role} that a user chooses for the data $i\in\I$, where
\begin{itemize}
\item $x=\textsc{s}$: a {sensor} for the data;
\item $x=\textsc{r}$: a requester for the data;
\item $x=\textsc{a}$: an {alien} for the data.
\end{itemize}
Note that when the user chooses to be a sensor or requester for the data, she needs to further select a quality $q_k$ for the data.
We denote $\pi_{vc}(x,q_k)$ as the \emph{payoff} of a type-$(v,c)$ user when choosing a role $x $ and a quality $q_k$. Note that users' decisions are coupled. We keep the notation $\pi_{vc}(x,q_k)$ for simplicity and present the detailed dependence relationship next.
The objective of the user is to make the proper decision on $x$ and $q_k$ to maximize her payoff.

Next we provide the formal definition for the user payoff $\pi_{vc}(x, q_k)$ under different choices of $x$ and  $q_k$.

\subsubsection{Sensor}

When choosing to be a sensor (i.e., $x = \textsc{s}$) with a quality $q_k$, the user can achieve a \emph{direct benefit} from the data based on (\ref{eq:value}) and (\ref{eq:cost}), i.e.,
 \begin{equation}
U_{vc}^{\textsc{s}}(q_k) = w\cdot v\cdot f(q_k)-c\cdot g(q_k),
 \end{equation}
where the first term denotes the user utility for the data, and the second term denotes the sensing cost for the data.

Moreover, the sensor can also share the data with requesters to get some sharing benefit.
Let $\Phi_k$ denote the \emph{average} sharing benefit of a sensor choosing quality $q_k$.
Then, the \emph{payoff} of a type-$(v,c)$ sensor choosing quality $q_k$ can be defined as:
 \begin{equation}
\label{eqsellerpayoff}
\begin{aligned}
 & \pi_{vc}(\textsc{s},q_k)=U_{vc}^{\textsc{s}}(q_k)+\Phi_k.
\end{aligned}
 \end{equation}
Obviously, the payoff of a sensor greatly depends on the average sharing benefit $\Phi_k$ that she can achieve. We will provide the detailed analysis for $\Phi_k$ in Sections \ref{sec:phi} and \ref{sec:phik}.

\subsubsection{Requester}

When choosing to be a requester (i.e., $x = \textsc{r}$) with a quality $q_k$,
the user obtains the data from a sensor who has sensed the data with quality $q_k$ already.
The requester can achieve a \emph{direct benefit} from the data:
 \begin{equation}
U_{vc}^{\textsc{r}}(q_k)=w \cdot v\cdot f(q_k)-\ctx,
 \end{equation}
where the first term denotes the user utility for the data, and the second term denotes the data transmission cost (including both the   uploading cost of the sensor and the   downloading cost of the requester) that the requester bears.

Moreover, the requester needs to provide some reward to the sensor, denoted as $\beta_{vc} (q_k )$. Based on the pricing scheme in (\ref{eq:price}), we have:
\begin{equation}\label{eq:payment}
\beta_{vc}(q_k) = U_{vc}^{\textsc{r}}(q_k) \cdot (1-\eta) + \wsp \cdot h(q_k),
\end{equation}
where the first term denotes the benefit that the requester shares with the sensor, and the second term denotes the quality-based price for the data.

Based on the above, the \emph{payoff} of a type-$(v,c)$ requester choosing quality $q_k$ can be defined as:\footnote{We have assumed that requesters cannot benefit from sharing, due to the time involved for requesters to obtain the data. That is, the value of the data will decrease after a requester obtains the data. This is reasonable due to the timeliness of the sensory data, which is very important in some mobile crowdsensing applications, e.g., the WiFi/cellular signal strength at a hotspot at a particular time \cite{opensignal}.}
\begin{align}
 & \pi_{vc}(\textsc{r},q_k)= U_{vc}^{\textsc{r}}(q_k) - \beta_{vc} (q_k).
\label{eqbuyerpayoff}
\end{align}

It is easy to see that a requester does not care about which sensor is sharing the data with her, due to the assumption of the identical data transmission cost among any pair of users.
In the case that there are multiple sensors holding the desired data of a requester,  the server will pick a sensor uniformly at random.

\subsubsection{Alien}

When choosing to be an alien (i.e., $x = \textsc{a}$),
the user neither senses the data nor requests the data from a sensor.
Thus, the \emph{payoff} of an alien is normalized to zero, i.e.,
 \begin{equation}\label{eqalienpayoff}
\pi_{vc}(\textsc{a})=0.
 \end{equation}

Notice that a user's payoff depends not only on her own choice but also on other users' choices. Each user optimizes her decisions on $x$ and $q_k$ to maximize her payoff, taking into account other users' decisions. Such a strategic interaction can be modeled by a non-cooperative game. Next, we present the detailed game formulation.

\subsection{Game Formulation}

We model the interactions of users (for the data $i\in\I$) as a non-cooperative game, called the CSRS (\underline{C}rowd\underline{S}ensing \underline{R}ole \underline{S}election) game. Specifically, the CSRS game consists of
\begin{itemize}

\item \emph{Players}: A set $\N = \{1,\cdots, N\}$ of mobile users, each associated with a type-$(v,c)$;

\item \emph{Strategies}: A set of $2K+1$ available choices
$\{(\textsc{s},q_k),(\textsc{r},q_k), \textsc{a}, \forall k\in\mathcal{K}\}$ for each player;

\item \emph{Payoffs}: The user payoffs under different choices are defined in (\ref{eqsellerpayoff}), (\ref{eqbuyerpayoff}), and (\ref{eqalienpayoff}).

\end{itemize}

For analytical convenience, we assume that the number of users $N$ is very large, such that the impact of a single user's choice on the whole population can be ignored. This assumption is mainly used for obtaining the closed-form result, and is also referred to as the ``non-atomic user'' assumption in the literature \cite{WardropNash}. The non-atomic user game provides the asymptotic result that often well approximates a practical system even with not so large number of users. The related solution concept is often called Wardrop equilibrium\cite{WardropNash}, which is usually easier to compute than Nash equilibrium, yet is a good approximation for Nash equilibrium\cite{WardropNash}. We will henceforth focus on the concept of Wardrop equilibrium in this paper.

In the following, we will first study the quality-unaware CSRS game in Section~\ref{sec_MarketEvolutionAnalysis}, where each data is associated with a single quality.
Then, we will further study the general scenario of quality-aware CSRS game in Section~\ref{sec:Quality}, where each data has different versions of different qualities.


\section{Game Equilibrium Analysis}\label{sec_MarketEvolutionAnalysis}

In this section, we study the equilibrium of the CSRS game in the scenario without data quality-awareness,
where each data is associated with a fixed quality $q$. As a result, we omit the quality index $k$ in Section~\ref{sec_MarketEvolutionAnalysis}.
Hence, each user has three choices: acting as a sensor or requester (with the given quality $q$), or acting as an alien.

In this case, the data value $v\cdot f(q )$ in \eqref{eq:value} and the sensing cost $c\cdot g(q )$ in \eqref{eq:cost} are both constants under a fixed user type-$(v,c)$.
The quality-based price $\wsp \cdot h(q)$ in \eqref{eq:price} is also a constant.
Without loss of generality, we normalize $f(q )=g(q )=h(q)=1$. Hence,  the data value is $v$, the sensing cost is $c$, and the quality-based price is $p$.
Furthermore,  the user payoffs defined in (\ref{eqsellerpayoff}) and  (\ref{eqbuyerpayoff}) can be rewritten as:
\begin{align}
\pi_{vc}(\textsc{s}) & =  w \cdot v - c + \Phi ,
\\
\pi_{vc}(\textsc{r} ) & =  \eta \cdot ( w\cdot v -\ctx )- p.
\end{align}
In addition, the reward defined in \eqref{eq:payment} can be rewritten as:
\begin{equation} \label{eq:payment2}
 \beta_{vc}= (1-\eta) (w\cdot v -\ctx ) + p.
\end{equation}
Accordingly, the user type can be equivalently defined as the data value $v$ and the sensing cost $c$, also denoted by $(v,c)$.
Moreover, the strategy of a type-$(v,c)$ user can be written as   $x(v,c) \in \{ \textsc{s},\textsc{r},\textsc{a} \}$, as the quality choice is fixed at $q$.\footnote{Here, we focus on deriving the symmetric equilibrium, where the users with the same type will always choose the strategy.}

We start from analyzing the best response for each user under a particular market state, which defines the market shares of different user roles (i.e., sensors, requesters, and aliens).
Then, we further capture the stable market shares that correspond to the game equilibrium.

In the following, we first provide the formal definition for the game equilibrium with   undifferentiated data quality.

\begin{definition}[Game Equilibrium]
A strategy profile $\{ x^{\ast}(v,c), \forall v,c\}$ is an equilibrium of the game,  if and only if
$$
\pi_{vc}(x^{\ast}(v,c))\geq \pi_{vc}(x),\quad \forall x \in\{\textsc{s},\textsc{r},\textsc{a}\},
$$
for the whole user population with any type-$(v,c)$.
\end{definition}

Given a strategy profile $\{ x (v,c), \forall v,c\}$, the market will be partitioned into three parts, each corresponding to one   choice of role in $ \{\textsc{s},\textsc{r},\textsc{a}\}$, which we call the \emph{market state}.
Let market shares $(\etase,\etaex,\etanl)$ denote the set of users choosing to be sensors, requesters, and aliens.
Then, we have $\etase  = \{(v,c): x(v,c)=\textsc{s}\}$, $\etaex  =  \{(v,c): x(v,c)=\textsc{r}\}$, and  $\etanl  =  \{(v,c): x(v,c)=\textsc{a}\}$.

In what follows, we will first derive the average sharing benefit of a sensor, i.e., $\Phi$.
Then, we will analyze the user best response and characterize the game equilibrium.

\subsection{Derivation of Average Sharing Benefit -- $\Phi$}\label{sec:phi}
Recall that the sharing benefit provided by a requester with type-$(v,c)$ is $\beta_{vc} (q) $ defined in \eqref{eq:payment2}.
Thus, given market shares $(\etase,\etaex,\etanl)$, the total sharing benefit provided by all requesters in $\etaex $ can be computed by
\begin{align}
B^{\textsc{se}} =& \textstyle  N\iint_{\etaex } \beta_{vc} (q) \cdot \zeta_{vc}(v,c)\mathrm{d}v\mathrm{d}c.\label{eq:sensingbenefit}
\end{align}
Furthermore, the total number of sensors in $\etase $ is
\begin{equation}\label{eq:sensinguser}
\textstyle  N^{\textsc{se}} =N\iint_{\etase }\zeta_{vc}(v,c)\mathrm{d}v\mathrm{d}c.
\end{equation}

As mentioned previously, all requesters' data requests will be distributed among all sensors (with the desired data) randomly and uniformly. Thus, the \emph{average} sharing benefit $\Phi$ that each sensor can achieve is
\begin{equation}\label{eqmarketeq}
\textstyle      {\Phi}=\frac{B^{\textsc{se}} }{N^{\textsc{se}} }
   = \frac{\iint_{\etaex } [(1-\eta)(w\cdot v-\ctx)+\wsp] \cdot \zeta_{vc}(v,c)\mathrm{d}v\mathrm{d}c}{\iint_{\etase }\zeta_{vc}(v,c)\mathrm{d}v\mathrm{d}c}.
\end{equation}

\subsection{Users' Best Responses}\label{sec:BestResponse}

Now we show how users update their actions based on the best responses, under an existing market share distribution $(\etase,\etaex,\etanl)$. This shows how the market evolves starting from any initial market shares. Later in Subsection \ref{sec:dynamicpricing}, we will propose a dynamic system that converges to the equilibrium.

A type-$(v,c)$ user will choose to be a sensor (i.e., $x(v,c)=\textsc{s}$), if her payoff as a sensor is higher than that as a requester or alien, i.e.,
$
\pi_{vc}(\textsc{s})> \max( \pi_{vc}(\textsc{a}),  \pi_{vc}(\textsc{r}) )
$.\footnote{We ignore the equality case, as the probability of having equalities is zero under the continuous distribution.}
This leads to
$$
\textstyle   v>\max\left(\frac{c-\Phi}{w},\frac{c-\Phi-\eta \ctx-\wsp}{w(1-\eta)}\right).
$$
Hence, the newly derived market share of sensors is
\begin{equation}
\textstyle    \widetilde{\etase} =\left\{(v,c): v>\max   \left(\frac{c-\Phi}{w},\frac{c-\Phi-\eta \ctx-\wsp}{w(1-\eta)}   \right) \right\}.
\end{equation}

A type-$(v,c)$ user will choose to be a requester (i.e., $x(v,c)=\textsc{r}$), if
$
\pi_{vc}(\textsc{r})> \max(\pi_{vc}(\textsc{a}),  \pi_{vc}(\textsc{s}) )
$.
This leads to
$$
\textstyle   \frac{\eta \ctx + \wsp }{w \eta} < v < \frac{c-\Phi-\eta \ctx-\wsp}{w(1-\eta)}.
$$
That is, the newly derived market share of requesters  is
\begin{equation}
\textstyle   \widetilde{\etaex} =\left\{(v,c): \frac{\eta \ctx + \wsp }{w \eta} < v < \frac{c-\Phi-\eta \ctx-\wsp}{w(1-\eta)} \right\}.
\end{equation}


A type-$(v,c)$ user will choose to be an alien (i.e., $x(v,c)=\textsc{a}$), if
$
\textstyle   \pi_{vc}(\textsc{a})> \max( \pi_{vc}(\textsc{s}),  \pi_{vc}(\textsc{r}))
$.
This leads to
$$
\textstyle   v < \min\left( \frac{c-\Phi}{w},\frac{\eta\ctx+ \wsp }{w \eta} \right).
$$
That is, the newly derived market share of aliens is
\begin{equation}
\textstyle    \widetilde{\etanl} =\left\{(v,c): v < \min  \left( \frac{c-\Phi}{w},\frac{\eta\ctx+ \wsp }{w \eta} \right) \right\}.
\end{equation}

\begin{figure}[t]
\begin{center}
\subfigure[Case 1]{
\begin{tikzpicture}[scale=1.07]
    \draw [<->,thick] (0,3) node (yaxis) [above] {$v$}
        |- (3,0) node (xaxis) [right] {$c$};
    \draw (0,2.5) coordinate (b_1) node () [left] {$1$} -- (2.5,2.5) coordinate (b_2)-- (2.5,0) coordinate (b_3) node () [below] {$1$};
    \draw [thick] (0.5,0) coordinate (b_4) -- (2.5,1.74) coordinate (b_5)  node () [right] {$l_1$};
    \draw [thick](1,0) coordinate (b_6) -- (2.5,2.3) coordinate (b_7) node () [right] {$l_2$};
    \draw [thick](0,1) coordinate (b_8) -- (2.5,1) coordinate (b_9) node () [right] {$l_3$};
    \coordinate (c) at (intersection of b_4--b_5 and b_8--b_9);
    \path[draw,thick,fill=white!20](c)--(b_9)--(b_7)--cycle;
    \path[draw,thick,fill=black!60](c)--(b_9)--(b_3)--(b_4)--cycle;
    \path[draw,thick,fill=black!10](c)--(b_4)--(0,0)--(b_1)--(b_2)--(b_7)--cycle;
    \draw [thick](0.5,0) coordinate (b_4) -- (2.5,1.74) coordinate (b_5)  node () [right] {$l_1$};
    \draw [thick](1,0) coordinate (b_6) -- (2.5,2.3) coordinate (b_7) node () [right] {$l_2$};
    \draw [thick](0,1) coordinate (b_8) -- (2.5,1) coordinate (b_9) node () [right] {$l_3$};

    \draw[dashed] (yaxis |- c) node[left] {}
        -| (xaxis -| c) node[below] {$\ctx+\frac{\wsp}{\eta}+\Phi$};
    \draw[dashed] (yaxis |- c) node[left] {}
        -| (xaxis -| c) node[below] {$\ctx+\frac{\wsp}{\eta}+\Phi$};
    \draw (1,1.5) node () {$\etase$};
    \draw (2.3,1.3) node () {$\etaex$};
    \draw (2,0.5) node () {$\etanl$};
    \fill[red] (c) circle (2pt);

            node[pos=0.5, auto=left] {\(\)};
\end{tikzpicture}
}
\subfigure[Case 2]{
\begin{tikzpicture}[scale=1.07]
    \draw [<->,thick] (0,3) node (yaxis) [above] {$v$}
        |- (3,0) node (xaxis) [right] {$c$};
    \draw (0,2.5) coordinate (b_1) node () [left] {$1$} -- (2.5,2.5) coordinate (b_2)-- (2.5,0) coordinate (b_3) node () [below] {$1$};
    \draw [thick] (0.5,0) coordinate (b_4) -- (2.5,2) coordinate (b_5)  node () [right] {$l_1$};
    \draw [thick] (1,0) coordinate (b_6) -- (2.25,2.5) coordinate (b_7) node () [above] {$l_2$};
    \draw [thick] (0,1) coordinate (b_8) -- (2.5,1) coordinate (b_9) node () [right] {$l_3$};
    \coordinate (c) at (intersection of b_4--b_5 and b_6--b_7);
    \coordinate (d) at (intersection of b_1--b_2 and b_6--b_7);
    \path[draw,thick,fill=white!20](c)--(b_9)--(b_2)--(d)--cycle;
    \path[draw,thick,fill=black!60](c)--(b_9)--(b_3)--(b_4)--cycle;
    \path[draw,thick,fill=black!10](c)--(b_4)--(0,0)--(b_1)--(d)--cycle;
    \draw [thick] (0.5,0) coordinate (b_4) -- (2.5,2) coordinate (b_5)  node () [right] {$l_1$};
    \draw [thick] (1,0) coordinate (b_6) -- (2.25,2.5) coordinate (b_7) node () [above] {$l_2$};
    \draw [thick] (0,1) coordinate (b_8) -- (2.5,1) coordinate (b_9) node () [right] {$l_3$};

    \draw[dashed] (yaxis |- c) node[left] {}
        -| (xaxis -| c) node[below] {$\ctx+\frac{\wsp}{\eta}+\Phi$};
    \draw[dashed] (yaxis |- c) node[left] {}
        -| (xaxis -| c) node[below] {$\ctx+\frac{\wsp}{\eta}+\Phi$};
    \draw (1,1.5) node () {$\etase$};
    \draw (2.3,1.5) node () {$\etaex$};
    \draw (2,0.5) node () {$\etanl$};
     \fill[red] (c) circle (2pt);
\end{tikzpicture}
}
\caption{Illustrations of lines $l_1$, $l_2$, and $l_3$. Gray region: Sensors; White region: Requesters; Black region: Aliens.}\label{comninecase2}
\end{center}
\end{figure}
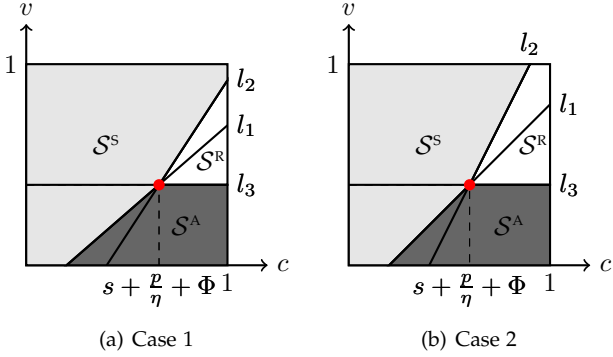

Based on the above discussion, we can see that the newly derived market shares $(\widetilde{\etase},\widetilde{\etaex},\widetilde{\etanl})$ can be characterized by the lines $l_1,l_2$, and $l_3$ illustrated in Fig.~\ref{comninecase2}, where
\begin{align}
l_1: & \textstyle   \quad v=\frac{c-\Phi}{w},\notag
\\
l_2: &  \textstyle   \quad v= \frac{c-\Phi-\eta \ctx-\wsp}{w(1-\eta)},\notag
\\
l_3: & \textstyle   \quad  v= \frac{\eta\ctx+ \wsp }{w \eta}.
\end{align}
Specifically,  the users with types above both lines $l_1$ and $l_2$ will choose to be sensors (gray region),
the users with types below line $l_2$ and above line $l_3$ will choose to be requesters (white region),
and the remaining users below line $l_1$ and line $l_3$ will choose to be aliens (black region). Furthermore, the three lines $l_1$, $l_2$, $l_3$ may intersect in two different ways, depending on the value of average sharing benefit $\Phi$.

Moreover, under the newly derived market shares $(\widetilde{\etase},\widetilde{\etaex},\widetilde{\etanl})$, the new average sharing benefit $\widetilde{\Phi} $ for each sensor can be updated using (\ref{eqmarketeq}), that is,
\begin{equation}\label{eqmarketeq-new}
\textstyle      \widetilde{\Phi}=  \frac{\iint_{\widetilde{\etaex} } [(1-\eta)(w\cdot v-\ctx)+\wsp] \cdot \zeta_{vc}(v,c)\mathrm{d}v\mathrm{d}c}{\iint_{\widetilde{\etase} }\zeta_{vc}(v,c)\mathrm{d}v\mathrm{d}c}.
\end{equation}
Note that $\widetilde{\Phi} $ is a function of the original average sharing benefit $\Phi$ (as the newly derived market shares $(\widetilde{\etase},\widetilde{\etaex},\widetilde{\etanl})$ are functions of $\Phi$), hence can be written as $\widetilde{\Phi} = \varphi (\Phi)$.
For convenience, we will write the newly derived $ {B^{\textsc{se}} }$ and $ {N^{\textsc{se}} }$ as functions of  $\Phi$ as well, i.e., $ {B^{\textsc{se}} }(\Phi)$ and $ {N^{\textsc{se}} } (\Phi)$. Then, we have
$\widetilde{\Phi} = \widetilde{\Phi} (\Phi) \triangleq \frac{{B^{\textsc{se}} }(\Phi)} {{N^{\textsc{se}} } (\Phi)}$.

\subsection{Game Equilibrium Analysis}\label{subsec:marketeq}
If a  strategy profile is an equilibrium, then none of the users has the incentive to change her strategy, which implies that the market shares and the average sharing benefit will no longer change.
This leads to the following necessary condition for the equilibrium.

\begin{proposition} \label{propequilibrium}
If a strategy profile $\{ x^{\ast}(v,c) ,  \forall v,c \}$ is an equilibrium, then the average sharing benefit $\Phi$ must satisfy the fixed-point condition:
$
\Phi =  \frac{B^{\textsc{se}}(\Phi)}{N^{\textsc{se}}(\Phi)}.
$
Furthermore, if $\Phi$ satisfies the fixed-point condition, the corresponding user decision profile $\{ x^{\ast}(v,c) \}$ is an equilibrium.
\end{proposition}

The proof can be found in the appendix.
Proposition~\ref{propequilibrium} implies that finding an equilibrium $\{ x^{\ast}(v,c), \forall v,c \}$ is equivalent to finding an equilibrium average sharing benefit $ {\Phi}^{\ast}$ that satisfies  $\Phi^{\ast}  =  \frac{B^{\textsc{se}}(\Phi^{\ast})}{N^{\textsc{se}}(\Phi^{\ast})}$.

Next, we analyze the existence and uniqueness of the equilibrium average sharing benefit $ {\Phi}^{\ast} $.
To solve
$
\Phi -\frac{B^{\textsc{se}}(\Phi )}{N^{\textsc{se}}(\Phi )}=0,
$
we can equivalently solve
\begin{equation}\label{eq:yyy}
\Phi \cdot N^{\textsc{se}}(\Phi )-B^{\textsc{se}}(\Phi )=0.
\end{equation}
For convenience, we define the following function:
\begin{equation}\label{eq:xxx}
\Lambda(\Phi) \triangleq \Phi \cdot N^{\textsc{se}}(\Phi)-B^{\textsc{se}}(\Phi).
\end{equation}
Hence, the problem of finding the equilibrium is equivalent to the problem of finding the roots of   $\Lambda(\Phi) = 0$.

As shown in Fig. \ref{comninecase2}, the three lines $l_1$, $l_2$, $l_3$ may intersect in two different ways, depending on the value of average sharing benefit $\Phi$.
For example, in the left subfigure, line $l_2$ intersects with the vertical boundary line $c=1$, while in the right subfigure, line $l_2$ intersects with the horizontal boundary line $v=1$.\footnote{It is possible that line $l_1$ may intersect with $v=1$ just like line~$l_2$. However, this case is equivalent to the case in the right subfigure, because they have the same partitions of $\etase$, $\etaex$, and $\etanl$, and the two intersection cases of line $l_1$ do not influence the boundary of $\etaex$.}
This will affect the computations of $ {B^{\textsc{se}}(\Phi) } $ and ${N^{\textsc{se}}(\Phi)}$, hence the computation of the root of $\Lambda(\Phi) = 0$.
Next, we analyze these two cases sequentially.

\subsubsection{Case 1: High Average Sharing Benefit}

When $\Phi$ is larger than a critical value $\Phi_0 $ defined below
\begin{equation}
\Phi_0 \triangleq 1-\eta \ctx-\wsp-w(1-\eta),
\end{equation}
 $l_2$ will intersect with the vertical boundary line $c=1$ (left subfigure in Fig. \ref{comninecase2}).
Intuitively, when $\Phi=\Phi_0$, $l_2$ will intersect at the cross point of lines $c=1$ and $v=1$ (i.e., the right upper corner).

In this case, the newly derived market shares  $(\widetilde{\etase},\widetilde{\etaex}, \widetilde{\etanl})$  can be derived according to the left subfigure in Fig. \ref{comninecase2}).
To avoid confusion with the low average sharing benefit case (i.e., case 2), we denote the corresponding functions $\Lambda(\Phi)$, $ N^{\textsc{se}}(\Phi)$,    $B^{\textsc{se}}(\Phi)$ by $\Lambda_h(\Phi)$, $B_{h}^{\textsc{se}}(\Phi)$,   $N_{h}^{\textsc{se}}(\Phi)$ in this case, which can be computed by \eqref{eq:sensinguser}, \eqref{eq:sensingbenefit}, and \eqref{eq:xxx}, respectively.\footnote{Detailed derivations and proofs can be referred to the appendix, unless otherwise mentioned.}

We can show that when $\Phi \geq \Phi_0$, function $\Lambda_h(\Phi) $ is monotonically increasing in $\Phi$ (see the appendix for details).
We further notice that $ \Lambda_h(\Phi) > 0 $ when $\Phi$ is large enough. Hence, the root of $ \Lambda_h(\Phi) =0 $ in the regime $\Phi \geq \Phi_0$ is determined by the value of $\Lambda_h(\Phi_0)$.
Formally, we have the following proposition.

\begin{proposition}\label{prop:uniquecase2general}
If $\Lambda_h(\Phi_0) < 0$, the quality-unaware CSRS game has a unique equilibrium with respect to $\Phi$ in the regime $[\Phi_0, +\infty)$;
  otherwise, it has no equilibrium in the regime $[\Phi_0, +\infty)$.
\end{proposition}

\subsubsection{Case 2: Low Average Sharing Benefit}

When $\Phi$ is smaller than the critical value
$ \Phi_0  $,
$l_2$ will intersect with the horizontal boundary line $v=1$ (right subfigure in Fig. \ref{comninecase2}).
In this case, the newly derived market shares  $(\widetilde{\etase},\widetilde{\etaex},\widetilde{\etanl})$  can be derived according to the right subfigure in Fig. \ref{comninecase2}.
To avoid confusion with case 1, we denote the corresponding functions $\Lambda(\Phi)$, $ N^{\textsc{se}}(\Phi)$, $B^{\textsc{se}}(\Phi)$ by $\Lambda_l(\Phi)$, $B_{l}^{\textsc{se}}(\Phi)$,  $N_{l}^{\textsc{se}}(\Phi)$  in this case, which can similarly be computed by  \eqref{eq:sensingbenefit}, \eqref{eq:sensinguser}, and \eqref{eq:xxx}, respectively.

We can show that when $\Phi \leq \Phi_0$,
the function $ \Lambda_l(\Phi) $ is either monotonically increasing with $\Phi$, or first decreasing with $\Phi$ and then increasing with $\Phi$ (hence unimodal).
We further notice that $ \Lambda_l(0) < 0 $.
Hence, the root of $ \Lambda_l(\Phi) = 0 $ in the regime $\Phi \leq \Phi_0$ is determined by the value of $\Lambda_l(\Phi_0)$.
Formally, we have the following proposition.

\begin{proposition}\label{prop:uniquecase1general}
If $\Lambda_l(\Phi_0) > 0$,  the quality-unaware CSRS game has a unique equilibrium with respect to $\Phi$ in the regime $[0, \Phi_0]$; otherwise,
it has no equilibrium in the regime $[0, \Phi_0]$.
\end{proposition}

\subsubsection{Game Equilibrium}

Combining Propositions \ref{prop:uniquecase2general} and
\ref{prop:uniquecase1general} in both cases, and notice that
$ \Lambda_h(\Phi_0) =\Lambda_l(\Phi_0) $, we have the following theorem for the existence and uniqueness of the equilibrium.

\begin{theorem}[Existence and Uniqueness]
\label{theoremgeneral}
The quality-unaware CSRS game has a unique equilibrium given by  $\Lambda(\Phi) = 0$, and
\begin{itemize}
\item if $\Lambda(\Phi_0) < 0$, the equilibrium is located in $( \Phi_0, +\infty)$;
\item if $\Lambda(\Phi_0) > 0$, the equilibrium is located in $(0, \Phi_0)$;
\item if $\Lambda(\Phi_0) = 0$, the equilibrium is   $  \Phi_0$.
\end{itemize}
\end{theorem}

In the next subsection, we will further study how to reach the equilibrium dynamically.

\subsection{Best Response Iterative Algorithm}\label{sec:dynamicpricing}

In this subsection, we propose a best response iterative algorithm to reach the above game equilibrium.

To describe the best response iteration, we first define a \emph{virtual} time-slotted system with slots $t = 1, 2, \cdots$ (each with a sufficiently small time period),
and allow users to change their decisions in every time slot based on the newly derived market shares.
Let $\Phi(t)$ be the average sharing benefit at time slot $t$, and denote the corresponding $B^{\textsc{se}}(\Phi)$ and $N^{\textsc{se}}(\Phi)$ as $B^{\textsc{se}}(\Phi(t))$ and $N^{\textsc{se}}(\Phi(t))$, respectively. According to the analysis in Section~\ref{subsec:marketeq}, we have
\begin{equation}\label{eq:iteration}
\Phi(t + 1) =
\frac{B^{\textsc{se}}(\Phi(t))}{N^{\textsc{se}}(\Phi(t))} = \left\{
\begin{aligned}
&\textstyle \frac{B_{h}^{\textsc{se}}(\Phi(t))}{N_{h}^{\textsc{se}}(\Phi(t))}, \quad  \text{ if }  \Phi(t)\leq \Phi_0,\\
&\textstyle \frac{B_{l}^{\textsc{se}}(\Phi(t))}{N_{l}^{\textsc{se}}(\Phi(t))},\quad  \text{ if } \Phi(t) \geq \Phi_0   .
\end{aligned}
\right.
\end{equation}

However, under the pure best response iteration mentioned above, the system may not converge.
To this end, we propose a generalized distributed best response iterative algorithm, i.e., Algorithm \ref{algo:max0}, where each user updates her own decision with probability $1-\lambda$ in each time slot in a distributed manner.
Then, the dynamics of $\Phi$ is given by
\begin{equation}\label{eq:dynamics}
\textstyle   \Phi(t+1)=\lambda \cdot  \Phi(t ) + (1-\lambda) \cdot \frac{B^{\textsc{se}}(\Phi(t))}{N^{\textsc{se}}(\Phi(t))}.
\end{equation}
Clearly, $\lambda = 0$ corresponds to the pure best response dynamics, while $\lambda = 1$ corresponds to a fixed network without dynamics.

\begin{algorithm}[t]
\DontPrintSemicolon 
\KwIn{Initial market shares $(\etase,\etaex,\etanl)$}
\KwOut{$\Phi^{\ast}$}
Compute $\Phi^{(1)}$ by (\ref{eq:iteration}), and set $\lambda$ according to (\ref{eq:iterationstep})\;
Set $t=1$ and $\Phi^{(0)}\leq\Phi^{(1)}-\varepsilon$ \;
\While{$\left|\Phi^{(t)}-\Phi^{(t-1)}\right|\geq\varepsilon$} {
$t=t+1$\;
Update $\Phi^{(t)}$ according to (\ref{eq:dynamics})\;
 \vspace{-1mm}
}
$\Phi^{\ast} \gets \Phi^{(t)}$    \;
\Return{Game equilibrium $\Phi^{\ast}$}\;
\caption{Distributed Best Response Iteration I}
\label{algo:max0}
\end{algorithm}

Next, we show the convergence of the above generalized best response iteration.
For convenience, we define
\begin{equation}
\textstyle \psi_l (\Phi) = \frac{ \mathrm{d} \frac{B_l^{\textsc{se}}(\Phi)}{N_l^{\textsc{se}}(\Phi)}  }{ \mathrm{d} \Phi }
\mbox{~~~and~~~}
\psi_h (\Phi) = \frac{ \mathrm{d} \frac{B_h^{\textsc{se}}(\Phi)}{N_h^{\textsc{se}}(\Phi)}  }{ \mathrm{d} \Phi }.
\end{equation}
Then, we have the following proposition for convergence.

\begin{proposition}
The generalized best response iteration in (\ref{eq:dynamics}) converges to the unique equilibrium $ {\Phi}$ given in Theorem \ref{theoremgeneral}, if $\lambda $ is larger than a threshold $ \lambda_0$, where
\begin{equation}\label{eq:iterationstep}
\lambda_0 = \left\{
\begin{aligned}
&\textstyle \max\left( \frac{\psi_h(\Phi_0)-1}{\psi_h(\Phi_0)+1},  \ 0\right), \text{ if } \Lambda(\Phi_0) < 0,\\
&\textstyle \max\left( \frac{\psi_l(0)-1}{\psi_l(0)+1}, \  0\right), \text{ if } \Lambda(\Phi_0) > 0.
\end{aligned}
\right.
\end{equation}
\end{proposition}

Intuitively, a smooth enough  iteration (by setting a large  $\lambda$) will guarantee that the  iteration (\ref{eq:dynamics}) converges to the unique equilibrium, at the cost of a slower convergence.


\section{Quality-aware Game Equilibrium Analysis}\label{sec:Quality}
In this section, we will study the quality-aware CSRS game equilibrium, where each data is associated with a set of discrete qualities $\mathcal{Q} \triangleq \{q_k:k\in\mathcal{K}\}$.
Hence, each user can choose different qualities when acting as a data sensor or requester.
Here we focus on the case where a requester with a quality can only obtain data from a sensor with the same quality.
Later in Section \ref{sec:QualitySharing}, we will consider the more general cross-quality data sharing, where a requester can obtain data from a sensor who has the data of a higher quality than the requester needs.

To facilitate the analysis, we adopt a simple pricing scheme with $\eta = 1$, which corresponds to a pure quality-based pricing scheme, i.e., $\wsp \cdot h(q_k)$ for quality $q_k$.
Let $x(v,c) \in \{ \textsc{s},\textsc{r},\textsc{a} \}$
denote the role choice of a type-$(v,c)$ user, and  $q(v,c) \in \mathcal{Q} $ denote the quality choice of a type-$(v,c)$ user.
Then,  the quality-aware game equilibrium can be formally defined as follows.

\begin{definition}[Quality-aware Game Equilibrium]
A strategy profile $\{ (x^{\ast}(v,c),q^{\ast}(v,c)) ,\ \forall v,c\}$ is an equilibrium of the quality-aware CSRS game,  if and only if
$$
\pi_{vc}(x^{\ast}(v,c),q^{\ast}(v,c))\geq \pi_{vc}(x,q ), \ \forall x \in\{\textsc{s},\textsc{r},\textsc{a}\},\forall q \in\mathcal{Q},
$$
for the whole user population with any type-$(v,c)$.
\end{definition}

Given a strategy profile $\{ (x (v,c),q(v,c)) ,\ \forall v,c \}$, the market will be partitioned into $2K+1$ parts, each corresponding to a choice of role and quality, which we call the \emph{market state}.
Let $\etase_k$, $\etaex_k$, and $\etanl $ denote the set of users choosing to be sensors with quality $q_k$, requesters with quality $q_k$, and aliens,
 called the \emph{market shares} of sensors with quality $q_k$, requesters with quality $q_k$, and aliens,
respectively.
Then, we have $\etase_k = \{(v,c): x(v,c)=\textsc{s}, q(v,c)= q_k\}$, $\etaex_k =  \{(v,c): x(v,c)=\textsc{r}, q(v,c)= q_k\}$ and $\etanl  =  \{(v,c): x(v,c)=\textsc{a}\}$.

In what follows, we will first derive the average sharing benefit of a sensor, i.e., $\Phi_k$.
Then, we will analyze the user best response and characterize the game equilibrium.

\subsection{Derivation of Average Sharing Benefit -- $\Phi_k$}\label{sec:phik}

We now derive the average sharing benefit $\Phi_k $.
Similar as \eqref{eq:sensingbenefit},
given market shares $(\etase_k,\etaex_k,\etanl)$, the total sharing benefit provided by all requesters with quality $q_k$ is
\begin{equation}\label{eq:Qualitysensingbenefit}
\textstyle     B^{\textsc{se}}_k = N \iint_{\etaex_k} \wsp\cdot h(q_k) \cdot \zeta_{vc}(v,c)\mathrm{d}v\mathrm{d}c.
\end{equation}
Furthermore, the total number of sensors with quality $q_k$   is
\begin{equation}\label{eq:Qualitysensinguser}
\textstyle     N^{\textsc{se}}_k =N\iint_{\etase_k}\zeta_{vc}(v,c)\mathrm{d}v\mathrm{d}c.
\end{equation}

Thus, the \emph{average} sharing benefit $\Phi_k$ that each sensor with quality $q_k$ can achieve is
\begin{align}
   {\Phi}_k=\frac{B^{\textsc{se}}_k}{N^{\textsc{se}}_k}
   &\textstyle   = \frac{\iint_{\etaex_k} \wsp\cdot h(q_k)\cdot \zeta_{vc}(v,c)\mathrm{d}v\mathrm{d}c}{\iint_{\etase_k }\zeta_{vc}(v,c)\mathrm{d}v\mathrm{d}c}
=\wsp\cdot h(q_k)\cdot \frac{|\etaex_k|}{|\etase_k|},\label{Qualityeqmarketeq}
\end{align}
where $ |\etase_k| = \iint_{\etase_k}\zeta_{vc}(v,c)\mathrm{d}v\mathrm{d}c$ is the percentage of sensors with quality $q_k$,
and $ |\etaex_k| = \iint_{\etaex_k}\zeta_{vc}(v,c)\mathrm{d}v\mathrm{d}c$ is the percentage of requesters with quality $q_k$. Intuitively,  $|\etaex_k|/|\etase_k|$ is the average number of data requests assigned to each sensor, and $\wsp\cdot h(q_k)$ is the average sharing benefit from each data request (sharing).
Obviously,  $\Phi_k$ decreases with the number of sensors with quality $q_k$, and increases with the   number of requesters with quality $q_k$.
For notational convenience, we denote $\boldsymbol{\Phi} \triangleq ({\Phi}_k, \forall k\in\K)$ as the average sharing benefit vector for all $k\in\K$.

\subsection{Users' Best Choices}\label{sec:BestResponseHeter}

Now we show how users update their actions based on the best responses, under an existing market share distribution $(\etase_k ,\etaex_k , \etanl)$.

A type-$(v,c)$ user will choose to be a sensor with quality $q_k$ (i.e., $x(v,c)=\textsc{s}$ and $q(v,c)=q_k$), if and only if her payoff as a sensor with quality $q_k$ is higher than that in another role with any other quality, i.e.,
\begin{equation}\label{eq:ss}
\pi_{vc}(\textsc{s}, q_k)\geq \max\big(\pi_{vc}(\textsc{s},q_j),\pi_{vc}(\textsc{r},q_j),\pi_{vc}(\textsc{a}),\forall j\in\mathcal{K}\big).
\end{equation}

Similarly, a type-$(v,c)$ user will choose to be a requester with quality $q_k$ (i.e., $x(v,c)=\textsc{r}$ and $q(v,c)=q_k$),  if and only if
\begin{equation}\label{eq:rr}
\pi_{vc}(\textsc{r},q_k)\geq \max\big(\pi_{vc}(\textsc{r},q_j),\pi_{vc}(\textsc{s},q_j),\pi_{vc}(\textsc{a}),\forall j\in\mathcal{K}\big),
\end{equation}
and choose to be an alien (i.e., $x(v,c)=\textsc{a}$), if and only if
\begin{equation}\label{eq:aa}
\pi_{vc}(\textsc{a})> \max\big(\pi_{vc}(\textsc{s},q_j),  \pi_{vc}(\textsc{r},q_j),\forall j\in\mathcal{K}\big).
\end{equation}

According to the above conditions, we can obtain the best choice of any user with any type-$(v,c)$, hence
derive the newly derived market shares $(\widetilde{\etase_k},\widetilde{\etaex_k}, \widetilde{\etanl})$ accordingly. Based on which,  we can further derive the new average sharing benefit $\widetilde{\Phi}_k, \forall k\in\mathcal{K}$ as follows:
\begin{equation}
\textstyle      \widetilde{\Phi}_k  =\wsp\cdot h(q_k) \cdot \frac{|\widetilde{\etaex_k}|}{|\widetilde{\etase_k}|}.
\end{equation}
Note that $\widetilde{\Phi}_k $ is a function of the original average sharing benefit vector $\boldsymbol{\Phi}$ (as both $\widetilde{\etase_k}$ and $ \widetilde{\etaex_k}$ are functions of $\boldsymbol{\Phi}$), hence can be written as $\widetilde{\Phi}_k  = \widetilde{\Phi}_k (\boldsymbol{\Phi})$.

\subsection{Quality-aware Game Equilibrium Analysis}
Now we analyze the (possible) existence and uniqueness of the quality-aware game equilibrium.

We first notice that if a  strategy profile is an equilibrium, then none of the users has the incentive to unilaterally change her strategy. This implies that the market shares will no longer change, and hence the average sharing benefits will no longer change.
This leads to the following necessary conditions for the quality-aware game equilibrium.

\begin{proposition} \label{propQualityequilibrium}
If a strategy profile $\{( x^{\ast}(v,c),q^{\ast}(v,c) ) , \forall v,c \}$ is a quality-aware game equilibrium, then the average sharing benefits satisfy the fixed-point conditions:
\begin{equation}\label{eq:fixed-pointcondition2}
\textstyle   \Phi_k = \wsp\cdot h(q_k)\cdot \frac{|\widetilde{\etaex_k}(\boldsymbol{\Phi})|}{|\widetilde{\etase_k}(\boldsymbol{\Phi})|},\forall k\in\mathcal{K}.
\end{equation}
Furthermore, if (\ref{eq:fixed-pointcondition2}) holds, the corresponding use decision profile  $\{( x^{\ast}(v,c),q^{\ast}(v,c) )\}$ is an equilibrium.
\end{proposition}

Proposition \ref{propQualityequilibrium} implies that finding a quality-aware game equilibrium $\{ (x^{\ast}(v,c),q^{\ast}(v,c)) , \forall v,c \}$ is equivalent to finding a set of equilibrium average sharing benefits $ {\Phi}_k^{\ast}, \forall k \in \mathcal{K},$ that satisfy  $\Phi_k^{\ast}= \wsp\cdot h(q_k)\cdot\frac{|\widetilde{\etaex_k}(\boldsymbol{\Phi})|}{|\widetilde{\etase_k}(\boldsymbol{\Phi})|}$ for all $ k \in \mathcal{K} $. This can be formally characterized by the following function set:
\begin{equation}
\label{eq:systemofequations}
\begin{bmatrix}
\Phi_1 \\
\Phi_2 \\
\vdots\\
\Phi_K
\end{bmatrix}
=
\begin{bmatrix}
\wsp\cdot h(q_1)\cdot \frac{|\widetilde{\etaex_1}|}{|\widetilde{\etase_1}|}  \\
\wsp\cdot h(q_2) \cdot \frac{|\widetilde{\etaex_2}|}{|\widetilde{\etase_2}|}
\\
\vdots
\\
\wsp\cdot h(q_K) \cdot \frac{|\widetilde{\etaex_K}|}{|\widetilde{\etase_K}|}
\end{bmatrix}
,
\end{equation}
where $\widetilde{\etaex_k}$ and $\widetilde{\etase_k}$, $\forall k\in \mathcal{K}$, are all functions of $\boldsymbol{\Phi} = (\Phi_k, \forall k\in\K)$, and can be derived according to the user best response analysis given in Section \ref{sec:BestResponseHeter}.

Next, we discuss the existence and uniqueness of the equilibrium average sharing benefits $\Phi_k^{\ast}, \forall k \in \mathcal{K}$.
Based on the above discussion, we can derive the equilibrium average sharing benefits $\Phi_k^{\ast}, \forall k \in \mathcal{K}$ by solving \eqref{eq:systemofequations}, or equivalently,
\begin{equation}
\Phi_k \cdot  |\widetilde{\etase_k}|-\wsp\cdot h(q_k)\cdot |\widetilde{\etaex_k}|=0.
\end{equation}
For convenience, we define the following functions:
\begin{equation}
\Lambda_k (\boldsymbol{\Phi} ) \triangleq \Phi_k \cdot |\widetilde{\etase_k}|-\wsp\cdot h(q_k) \cdot |\widetilde{\etaex_k}|, \ \forall k\in\mathcal{K}.
\end{equation}
Then, the problem of finding the equilibrium is equivalent to  finding the roots of $\Lambda_k(\boldsymbol{\Phi}) = 0, \forall k\in\mathcal{K}$.

Note that in the previous quality-unaware CSRS game, we only need to find a single variable $\Phi$ to solve \eqref{eq:yyy}, which can be analytically derived and efficiently solved by many classic methods such as the dichotomizing search.
When considering the quality-aware CSRS game here, however,
we need to compute $K$ variables $(\Phi_k, \forall k\in\mathcal{K})$ jointly to solve the function set~(\ref{eq:systemofequations}), and it is difficult to compute the fixed-point solutions $(\Phi_k^{\ast}, \forall k\in\mathcal{K})$ analytically due to the complicated coupling of $(\Phi_k, \forall k\in\mathcal{K})$, e.g., in the derivations of $\widetilde{\etaex_k} (\boldsymbol{\Phi})$ and $\widetilde{\etase_k}(\boldsymbol{\Phi})$.

In the next subsection, we will show that the best response iterative algorithm converges to the equilibrium dynamically, which implies the existence of equilibrium.

\subsection{Best Response Iterative Algorithm}
In this subsection, we design a distributed iterative algorithm based on the best response update to compute the quality-aware game equilibrium.
Algorithm \ref{algo:max} shows the detailed procedure of the proposed algorithm.
Specifically, in each round $t$, given the market shares $(\etase_k,\etaex_k,\etanl,\forall k\in\mathcal{K})$ in the previous round $t-1$,
the algorithm first computes the newly derived market shares by \eqref{eq:ss}-\eqref{eq:aa}. Then, it updates the average sharing benefits $(\Phi_k^{(t)},\forall k\in\K)$ according to \eqref{eq:algo}, where each user updates her own decision with probability $1-\lambda$ in each time slot in a distributed manner.
The above procedure repeats until the average sharing benefits do not change.

\begin{algorithm}[t]
\DontPrintSemicolon 
\KwIn{Initial market shares $(\etase_k,\etaex_k,\etanl:k\in\mathcal{K})$}
\KwOut{$(\Phi_k^{\ast},\forall k\in\K)$}
Compute $(\Phi_k^{(1)},\forall k\in\K)$ by (\ref{Qualityeqmarketeq}) based on the Input\;
Set $t=1$ and $\Phi_k^{(0)}\leq\Phi_k^{(1)}-\frac{\varepsilon}{K},\forall k\in\K$\;
\While{$\sum\limits_{k=1}^{K}\left|\Phi_k^{(t)}-\Phi_k^{(t-1)}\right|\geq\varepsilon$} {
  $t=t+1$\;
$
\etase_k \gets \{(v,c): \mbox{satisfying \eqref{eq:ss}}\},\ \forall k\in\mathcal{K}
$\;
$
\etaex_k \gets \{(v,c): \mbox{satisfying \eqref{eq:rr}}\},\ \forall k\in\mathcal{K}
$\;
$
\etanl \gets \{(v,c): \mbox{satisfying \eqref{eq:aa}}\}
$\;
Compute $(\Phi_k^{\dag},\forall k\in\K)$ by (\ref{Qualityeqmarketeq})\;
Update $(\Phi_k^{(t)},\forall k\in\K)$ according to:
 \begin{equation}\label{eq:algo}
\Phi_k^{(t)} = \lambda \Phi_k^{(t-1)}+(1-\lambda)\Phi_k^{\dag},\ \forall k\in\mathcal{K}
 \end{equation}
 \vspace{-5mm}
\;
}
 $\Phi_k^{\ast} \gets \Phi_k^{(t)},\ \forall k\in\mathcal{K}$\;
\Return{Quality-aware game equilibrium $(\Phi_k^{\ast},\forall k\in\K)$}\;
\caption{Distributed Best Response Iteration II}
\label{algo:max}
\end{algorithm}

It is easy to see that if Algorithm \ref{algo:max} converges, its converged state must be an equilibrium.
This implies that if we can prove the convergence of Algorithm \ref{algo:max}, then the existence of equilibrium can be guaranteed.
The following proposition shows the convergence of Algorithm \ref{algo:max}.
\begin{proposition}[Existence]\label{propalgorithm}
Algorithm \ref{algo:max} converges to an equilibrium  of the quality-aware CSRS game.
\end{proposition}

Note that it is difficult to prove the uniqueness of the quality-aware CSRS game equilibrium analytically, which is equivalent to the problem of determining the unique solution of \eqref{eq:systemofequations}.
Nevertheless, we can show that whenever the initial market shares $(\etase_k,\etaex_k,\etanl,\forall k\in\mathcal{K})$ are given, Algorithm \ref{algo:max} will converge to a unique equilibrium. We are not able to rule out the possibility that different choices of initial market shares lead to different equilibria. Furthermore, the observation from simulations in Section \ref{sec:NumericalResults} confirms the uniqueness of the equilibrium under a broad choice of the initial market shares.

\section{Game Equilibrium Analysis with Cross-Quality Data Sharing}\label{sec:QualitySharing}
In Section \ref{sec:Quality}, we have considered the scenario where a data sensor can only share her data to the requesters with the same quality requirement.
In this section, we consider a more general \emph{cross-quality} data sharing  scenario, where \emph{a sensor can also share her data with the requesters who have the lower quality requirements}.
This can be very useful in practice, as a high-quality data can often be transformed into a low-quality one (e.g., a high-resolution photo or video can be transformed into a low-resolution one through down sampling).
We will analyze how the cross-quality data sharing affects the user behaviors as well as the market equilibrium.

Specifically, when a high-quality data sensor shares her data with a low-quality data requester, the sensor will first transform the data into a low-quality one as requested, and then shares the transformed low-quality data with the requester.
Such a process is transparent to the requester, who does not need to pay additional fees because of such data transformation.

For the data sensors, however, the above cross-quality data sharing can have a significant impact. Specifically, for a sensor with quality $q_k$, her data can be potentially shared with more requesters, i.e., those with a quality requirement lower than $q_k$. Let $\Phi_{k,i}$ denote the \emph{average} sharing benefit that each sensor with quality $q_k$ can achieve from requesters with quality $q_i$, where $i \leq k$.
Note that each requester with quality $q_i$ can obtain data from any sensor with a quality no smaller than $q_i$, with a fixed payment $\wsp\cdot h(q_i)$.
Thus, we can derive $\Phi_{k,i}$ as follows:
\begin{equation}
\textstyle    \Phi_{k,i} = \wsp\cdot h(q_i) \cdot \frac{|\etaex_i|}{\sum_{j=i}^{K}|\etase_j|},
\end{equation}
where $|\etaex_i|$ is the percentage of requesters with quality $q_i$, and $\sum_{j=i}^{K}|\etase_j|$ is the total percentage of sensors with quality no smaller than $q_i$.
Thus, the overall \emph{average} sharing benefit $\Phi_k$ that each sensor with quality $q_k$ can achieve (from all requesters with quality $q_i \leq q_k$) is
\begin{equation}\label{eq:across}
\textstyle   \Phi_k = \sum\limits_{i=1}^k  \Phi_{k,i} =  \sum\limits_{i=1}^k  \frac{\wsp\cdot h(q_i) \cdot  |\etaex_i|}{\sum_{j=i}^{K}|\etase_j|}.
\end{equation}
By comparing \eqref{eq:across} with \eqref{Qualityeqmarketeq}, we can see two different impacts: (i) a sensor can potentially share data with more requesters (i.e., those requesters with a lower quality); (ii) a sensor will face more severe competition (i.e., from those sensors with a higher quality) for each request.

Similar to Proposition  \ref{propQualityequilibrium},
we have the following necessary conditions for the equilibrium with cross-quality sharing.

\begin{proposition} \label{propequilibrium2}
A strategy profile $\{( x^{\ast}(v,c),q^{\ast}(v,c) ) , \forall v,c \}$ is an equilibrium with cross-quality data sharing, if and only if the average sharing benefits $(\Phi_k,k\in\K)$ satisfy:
\begin{equation}\label{eq:equilibriumQualitySharing}
\textstyle \Phi_k =  \widetilde{\Phi}_k (\boldsymbol{\Phi}) \triangleq \sum\limits_{i=1}^k \frac{\wsp\cdot h(q_i)\cdot |\widetilde{\etaex_i}(\boldsymbol{\Phi})|}{\sum_{j=i}^{K}|\widetilde{\etase_j}(\boldsymbol{\Phi})|}, \ \forall k\in \mathcal{K},
\end{equation}
where $(\widetilde{\etase_j}(\boldsymbol{\Phi}),\forall j\in\{i,\cdots,K\})$ and $(\widetilde{\etaex_i}(\boldsymbol{\Phi}),\forall i\in\{1,\cdots,k\})$ are the newly derived market shares by \eqref{eq:ss}-\eqref{eq:aa}; $\widetilde{\Phi}_k (\boldsymbol{\Phi}) $ is the according new average sharing benefit.
\end{proposition}

We skip the detailed analysis for this scenario, as it is similar as that in Section \ref{sec:Quality}, except that we use the new formulation \eqref{eq:across} to compute the average sharing benefit.
We can design a similar best response iterative algorithm to reach the equilibrium, as shown in Algorithm \ref{algo:max2}.

\begin{algorithm}
\DontPrintSemicolon 
\SetAlgoLined
\KwIn{Initial market shares $(\etase_k,\etaex_k,\etanl:k\in\mathcal{K})$}
\KwOut{$(\Phi_k^{\ast},\forall k\in\K)$}
Compute $(\Phi_k^{(1)},\forall k\in\K)$ by (\ref{eq:across}) based on the Input\;
Set $t=1$ and $\Phi_k^{(0)}\leq\Phi_k^{(1)}-\frac{\varepsilon}{K},\forall k\in\K$\;
\While{$\sum\limits_{k=1}^{K}\left|\Phi_k^{(t)}-\Phi_k^{(t-1)}\right|\geq\varepsilon$} {
  $t=t+1$\;
$
\etase_k \gets \{(v,c): \mbox{satisfying \eqref{eq:ss}}\},\ \forall k\in\mathcal{K}
$\;
$
\etaex_k \gets \{(v,c): \mbox{satisfying \eqref{eq:rr}}\},\ \forall k\in\mathcal{K}
$\;
$
\etanl \gets \{(v,c): \mbox{satisfying \eqref{eq:aa}}\}
$\;
Compute $(\Phi_k^{\dag},\forall k\in\K)$ by  (\ref{eq:across}) \;
Update $(\Phi_k^{(t)},\forall k\in\K)$ according to (\ref{eq:algo}).
\;
}
 $\Phi_k^{\ast} \gets \Phi_k^{(t)},\ \forall k\in\mathcal{K}$\;
\Return{Cross-quality game equilibrium $(\Phi_k^{\ast},\forall k\in\K)$}\;
\caption{Distributed Best Response Iteration III}
\label{algo:max2}
\end{algorithm}

\begin{figure*}
\begin{minipage}[t]{0.32\textwidth}
\centering
\includegraphics[scale=0.3]{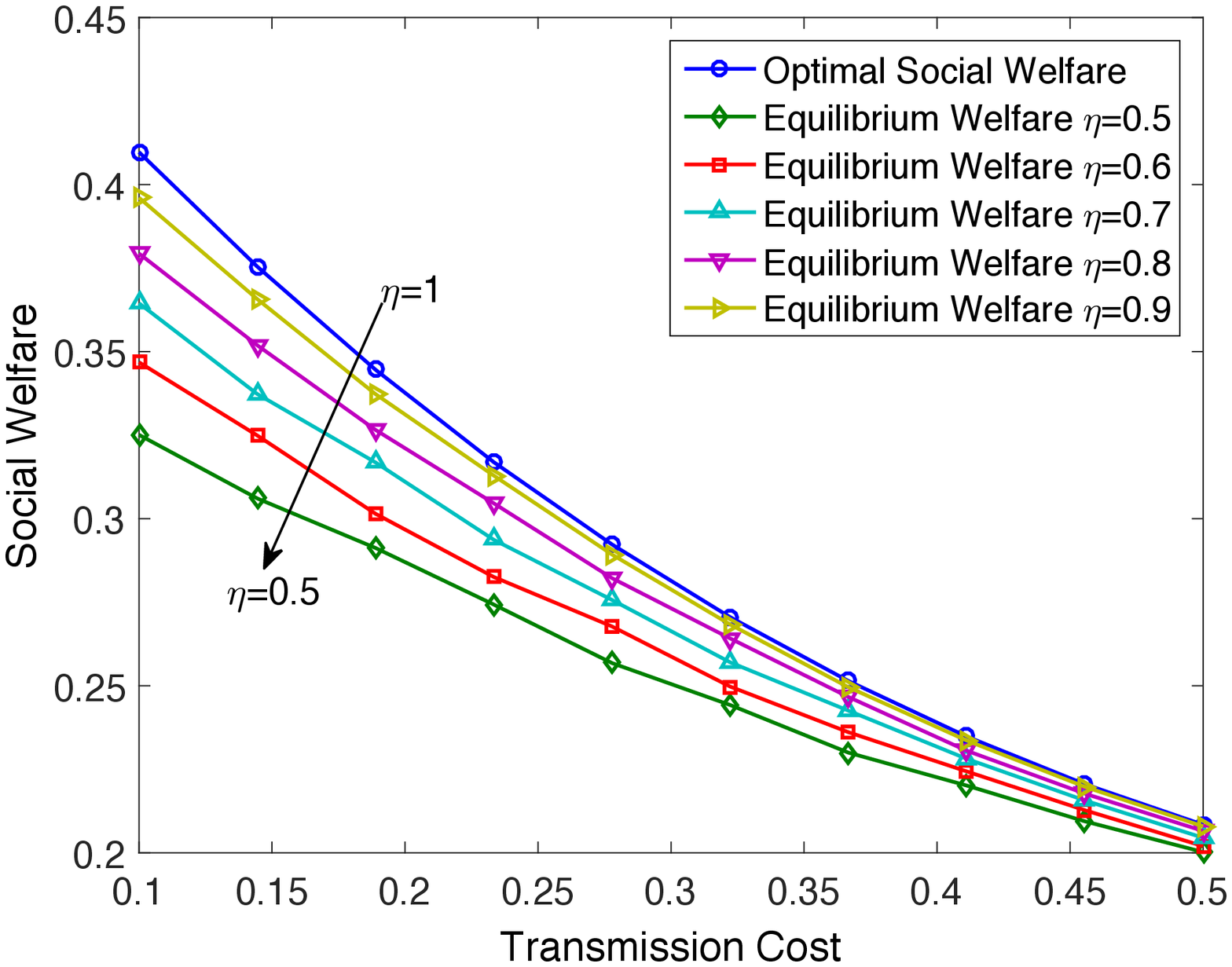}
\caption{Social welfare under different revenue sharing factors $\eta$ (price $\wsp=0$).}
\label{Fig:SocialWelfareDifferentParameters}
\end{minipage}%
\begin{minipage}[t]{0.02\textwidth}
~~~
\end{minipage}%
\begin{minipage}[t]{0.32\textwidth}
\centering
\includegraphics[scale=.3]{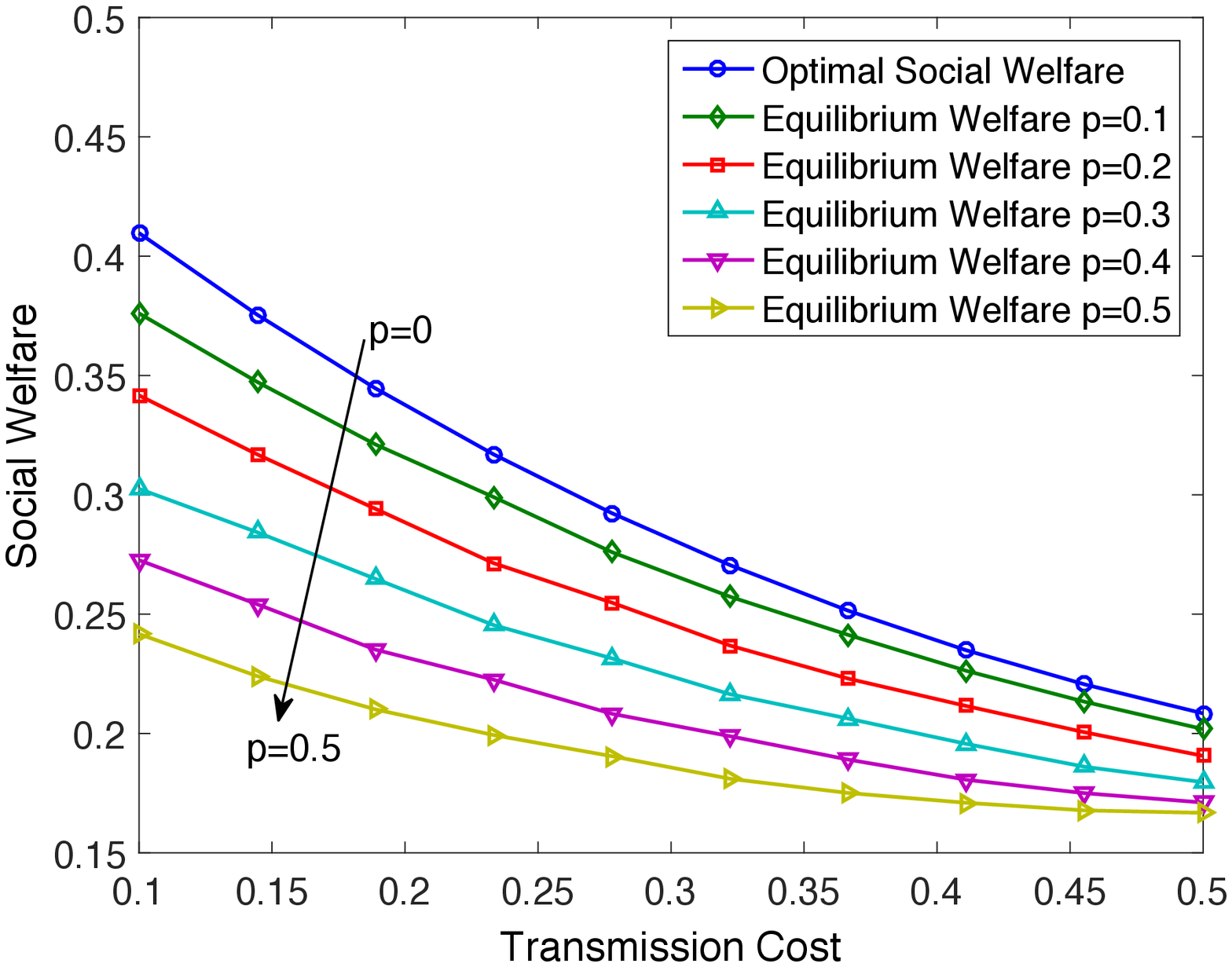}
\caption{Social welfare comparison under different prices $\wsp$ (revenue sharing factor $\eta=1$).}
\label{Fig:SocialWelfareWholesaleNash}
\end{minipage}%
\begin{minipage}[t]{0.02\textwidth}
~~~
\end{minipage}%
\begin{minipage}[t]{0.32\textwidth}
\centering
\includegraphics[scale=.3]{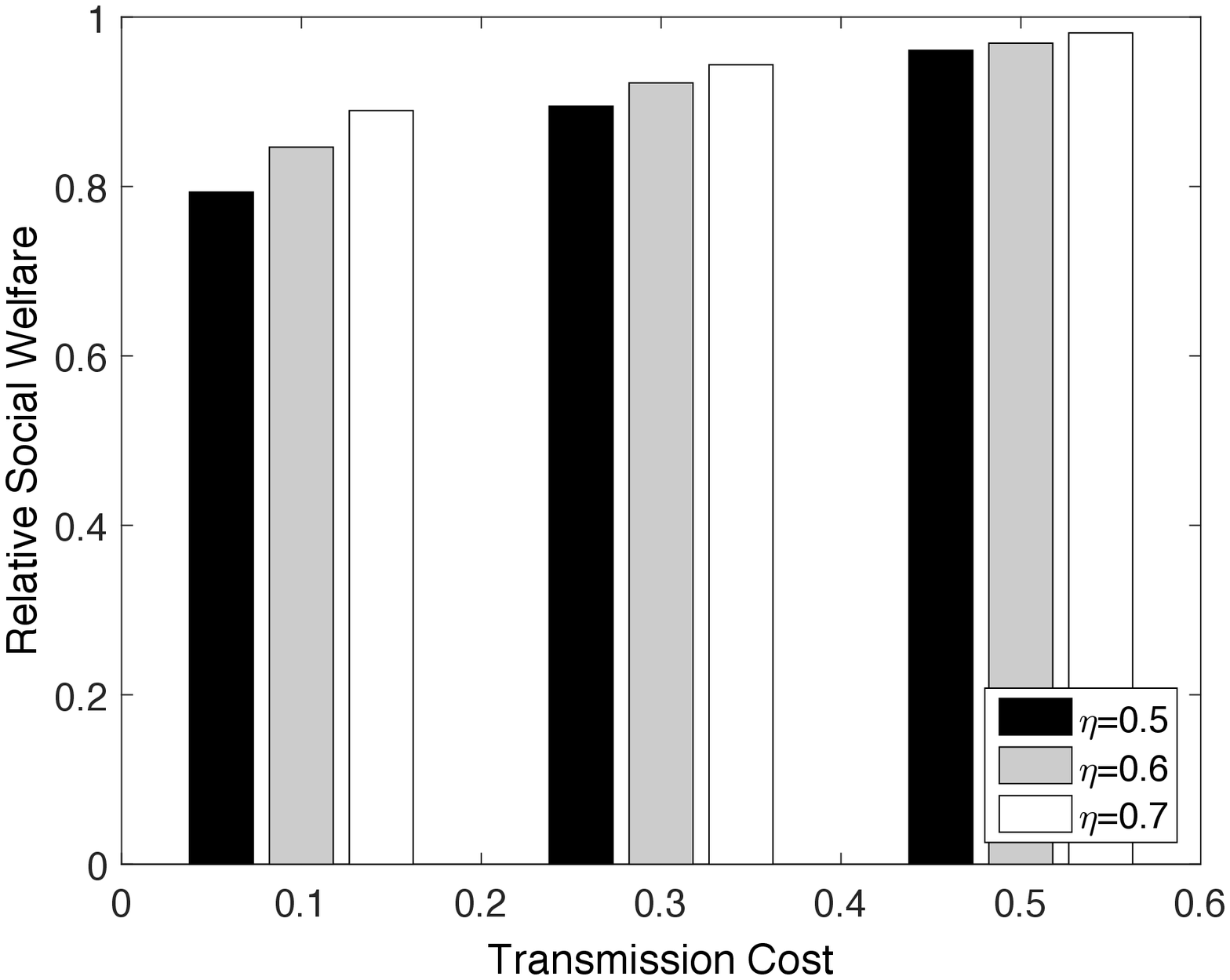}
\caption{Ratio of the welfare to the optimal social welfare under different factors $\eta$ (price $\wsp=0$).}
\label{Fig:RelativeWelfare}
\end{minipage}%
\end{figure*}

\section{Simulation and Evaluation}\label{sec:NumericalResults}

In this section, we provide the simulation results for the quality-unaware (CSRS) game and the quality-aware (CSRS) game, respectively.
In both scenarios, we will show the efficiency of the game equilibrium and the market shares under the game equilibrium.

\subsection{Simulation Setup}
We fix the number of users/devices as 1000, and randomly generate the (marginal) data value $v$ and the  (marginal) sensing cost $c$, both of which follow the uniform distribution on $[0,1]$.  In each simulation, we randomly generate 1000 systems (in terms of realizations of $v$ and $c$) and compute the average
outcome of all systems as the simulation result.

In the quality-unaware game, there are two parameters, i.e., the revenue sharing factor $\eta$ and the (fixed) price~$p$. To show the impacts on the equilibrium social welfare, we vary $\eta$ from  $\eta=0.5$ to  $\eta=1$ with an increment of 0.1, and vary $p$ from $p=0$ to $p=0.5$ with an increment of 0.1. We further vary  the transmission cost $s$  from 0.1 to 0.5. Furthermore, to show the  impacts on the equilibrium market shares, we choose three scenarios in terms of $\eta$, i.e., $\eta=0.4$ for small $\eta$, $\eta=0.8$ for large $\eta$,  and $\eta=1$ to fully remove the revenue sharing effect to the data sensors. In all three scenarios, we vary the transmission cost $s$ from 0 to 1.

In the quality-aware game, to characterize the impact of the quality-aware prices on the equilibrium social welfare, we choose two quality types ($K=2$), i.e., $q_1 =1$ as the low quality and $q_2=2$ as the high quality. The quality-aware data value from consuming the data is $0.4+v\log(1+q_k)$. The quality-aware sensing cost due to physical resources consumption is $0.1+cq_k^{0.5}$. The transmission cost $s $ varies in the range of $ [0.1, 0.5]$. We choose the quality-aware pricing scheme $p = 0.1+\{0.2, 0.5, 0.8\}\cdot q_k$, which correspond to ``small'', ``medium'', and ``large'' prices, respectively. Furthermore, to show the  impact of the quality-aware prices on the equilibrium market shares, we choose the two quality types, the quality-aware data value, and the quality-aware sensing cost similarly as those in the simulation of the equilibrium social welfare. We randomly generate the marginal data value $v$ and the marginal sensing cost $c$ with the uniform distributions on $[0,1]$. We choose the transmission cost $s = 0.2$ and the pricing scheme $p = 0.1+0.35q_k$.

Our key targets are to show the behaviors of users from the system level and the market partition due to users' choices. In particular, we want to understand the social benefit that can be achieved from the strategic data sensing and sharing among users. We consider that the system has a good performance if the equilibrium social welfare is close to the optimal social welfare benchmark.

\subsection{Simulations  of the Quality-unaware Game}\label{sec:numerical}
Now we provide simulation results to illustrate the efficiency of the quality-unaware game equilibrium, i.e., the ratio of the equilibrium social welfare to the maximum social welfare benchmark (in a centralized optimization). Here, social welfare is the sum of all users' payoffs. We will illustrate the absolute and relative equilibrium social welfares under different system parameters.
Moreover, we will also illustrate the equilibrium market shares under different system parameters.

\subsubsection{Maximum Social Welfare Benchmark}

The social welfare is defined as the sum of all users' payoffs. Hence, given a particular market partition $(\etase,\etaex,\etanl)$, the social welfare is
\begin{equation}\label{eq:SW}
\begin{aligned}
W& \textstyle  =\iint_{\etase}(v-c)\mathrm{d}v\mathrm{d}c+\iint_{\etaex}(v-\ctx)\mathrm{d}v\mathrm{d}c.\\
\end{aligned}
\end{equation}
where the social welfare generated by a sensor is $v-c$, and the social welfare generated by a requester  is $v-s$.

From a centralized optimization perspective (to maximize the social welfare), those users with type-$(v,c)$ that satisfies $v>c$ and $c<s$ should choose to be sensors, those users with type-$(v,c)$ that satisfies $v>s$ and $c>s$ should choose to be requesters, and all other users will choose to be aliens. Accordingly, we can compute the maximum social welfare benchmark. For more details, please refer to the appendix.

\begin{figure*}
\begin{minipage}[t]{0.32\textwidth}
\centering
\begin{overpic}[scale=0.3]{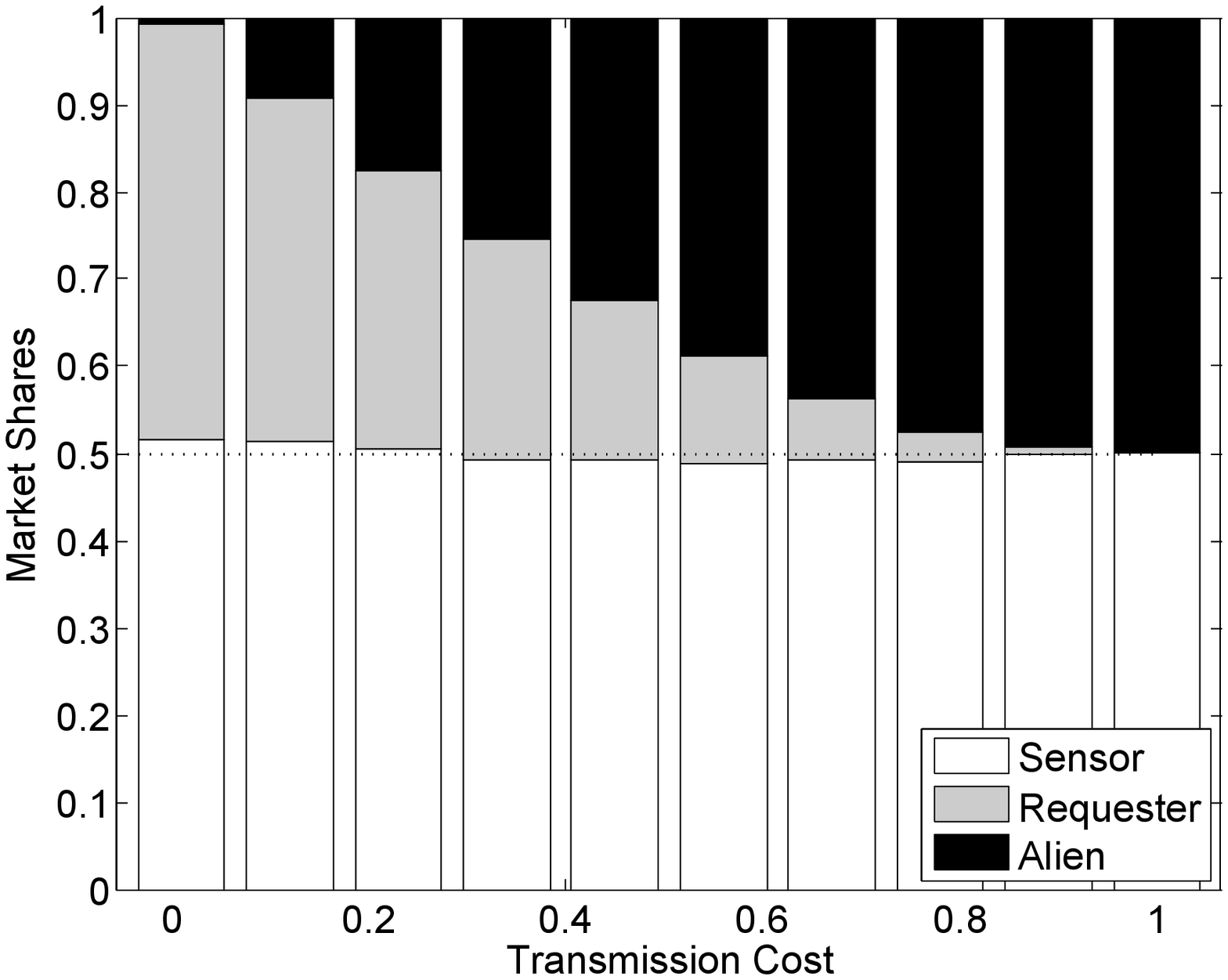}
\put(99,54){\rotatebox{90}{{\tiny Alien}}}
\thicklines\put(98,58){\rotatebox{0}{\color{gray}\vector(-1,0){8}}}

\put(9,79){\rotatebox{0}{{\tiny Requester}}}
\thicklines\put(15,78){\rotatebox{270}{\vector(1,0){8}}}

\put(99,21){\rotatebox{90}{{\tiny Sensor}}}
\thicklines\put (98,25){\rotatebox{0}{\vector(-1,0){8}}}
\end{overpic}
\caption{Market shares under the revenue sharing factor $\eta=0.4$ and the price $\wsp=0$.}
\label{Fig:NashBarMarketSharesSmallRho}
\end{minipage}%
\begin{minipage}[t]{0.02\textwidth}
~~~
\end{minipage}%
\begin{minipage}[t]{0.32\textwidth}
\centering
\begin{overpic}[scale=0.3]{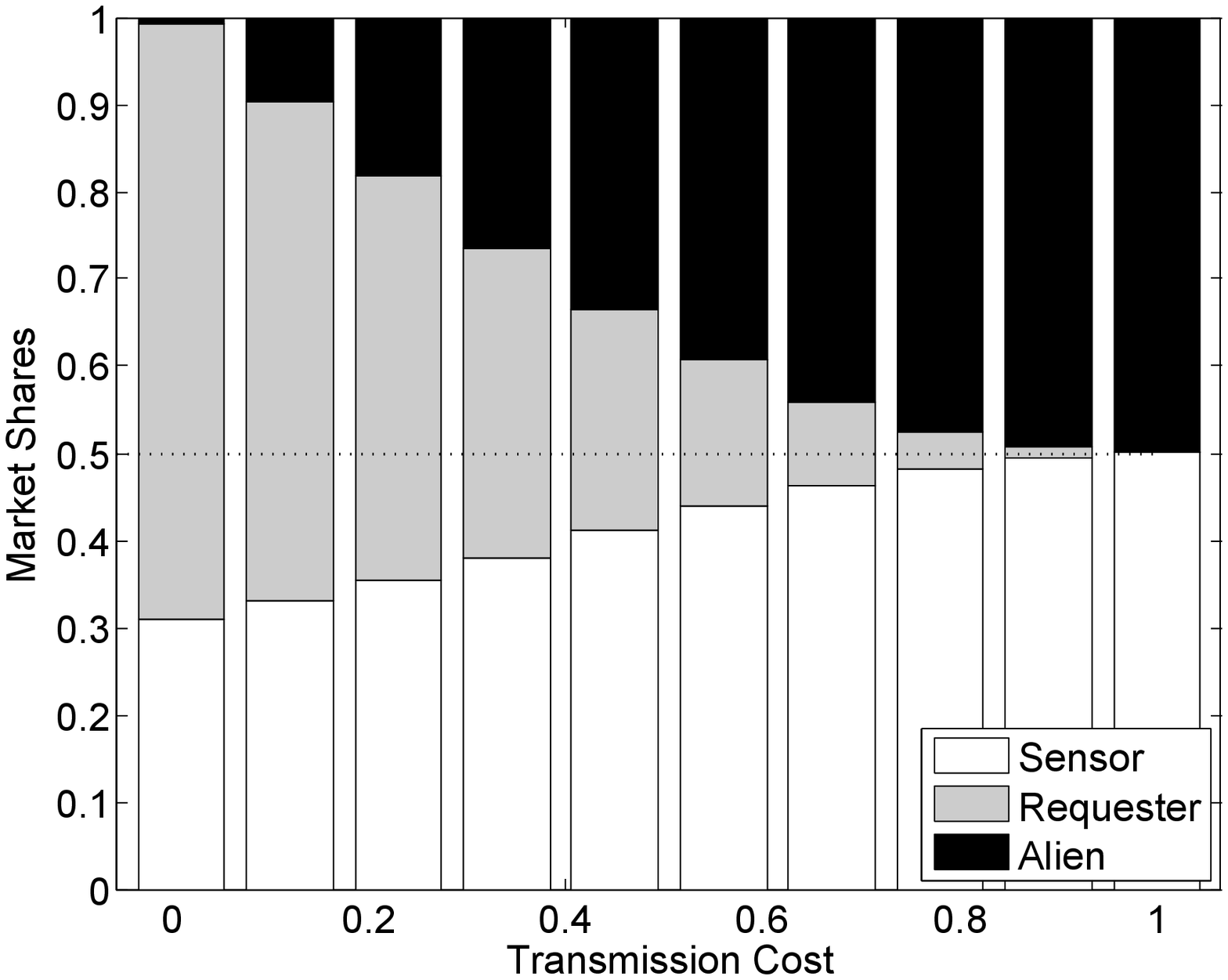}
\put(99,54){\rotatebox{90}{{\tiny Alien}}}
\thicklines\put(98,58){\rotatebox{0}{\color{gray}\vector(-1,0){8}}}

\put(9,79){\rotatebox{0}{{\tiny Requester}}}
\thicklines\put(15,78){\rotatebox{270}{\vector(1,0){8}}}

\put(99,21){\rotatebox{90}{{\tiny Sensor}}}
\thicklines\put (98,25){\rotatebox{0}{\vector(-1,0){8}}}
\end{overpic}
\caption{Market shares under the revenue sharing factor $\eta=0.8$ and the price $\wsp=0$.}
\label{Fig:NashBarMarketSharesLargeRho}
\end{minipage}%
\begin{minipage}[t]{0.02\textwidth}
~~~
\end{minipage}%
\begin{minipage}[t]{0.32\textwidth}
\centering
\begin{overpic}[scale=0.3]{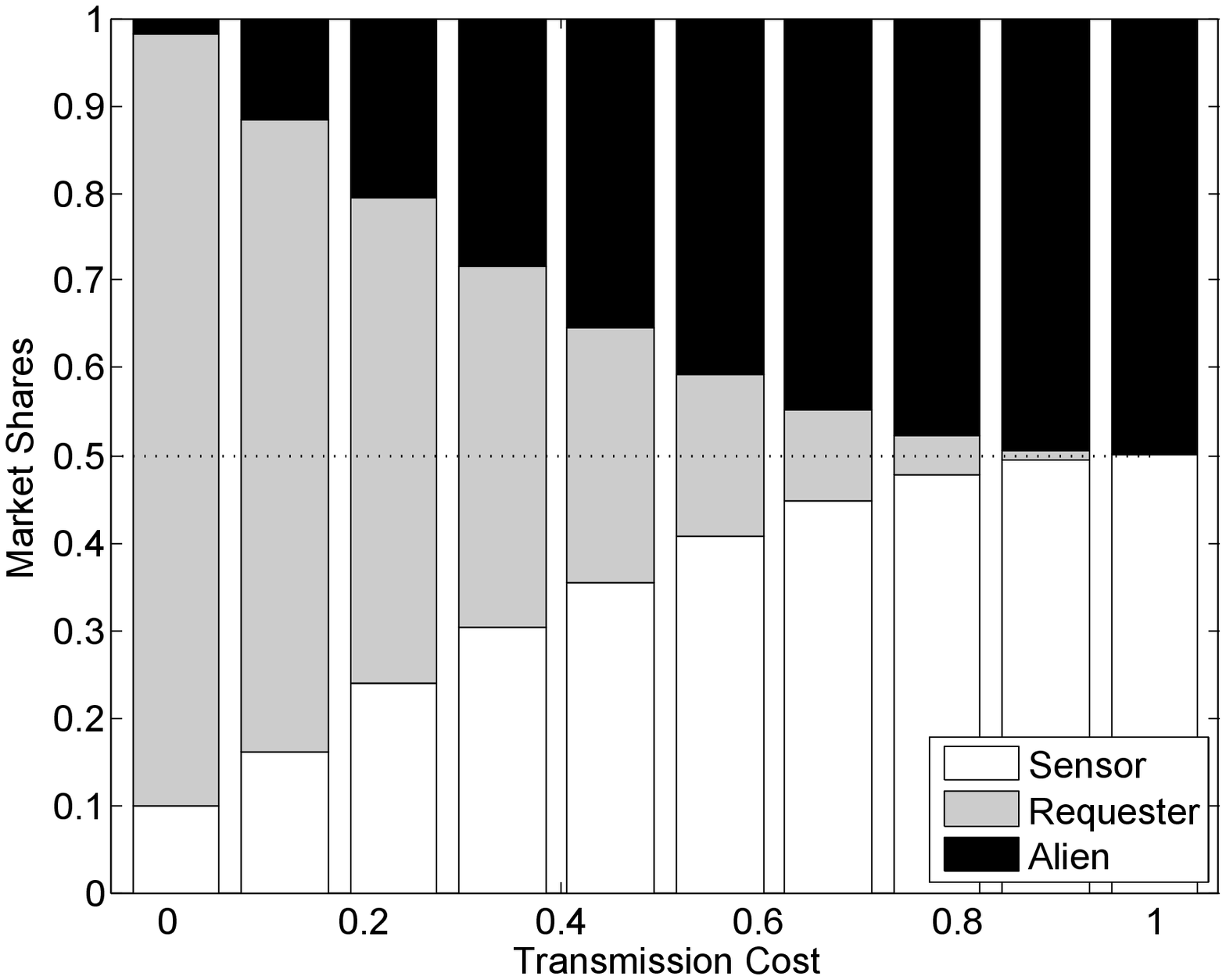}
\put(99,54){\rotatebox{90}{{\tiny Alien}}}
\thicklines\put(98,58){\rotatebox{0}{\color{gray}\vector(-1,0){8}}}

\put(8.5,79){\rotatebox{0}{{\tiny Requester}}}
\thicklines\put(14.5,78){\rotatebox{270}{\vector(1,0){8}}}

\put(99,21){\rotatebox{90}{{\tiny Sensor}}}
\thicklines\put (98,25){\rotatebox{0}{\vector(-1,0){8}}}
\end{overpic}
\caption{Market shares under the revenue sharing factor $\eta=1$ and the price $\wsp=0$.}
\label{Fig:WholesaleBarMarketShares}
\end{minipage}%
\end{figure*}

\begin{figure*}[t]
\centering
\subfigure[Social welfare under matching-quality vs. maximum social welfare benchmark]{\label{fig:9a}
\includegraphics[scale=0.29]{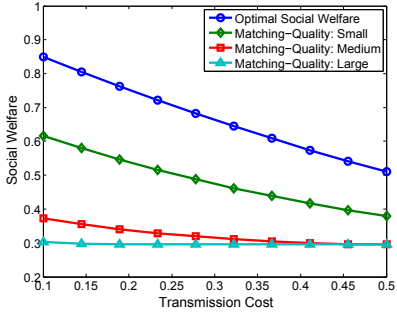}}
~
\subfigure[Social welfare under cross-quality vs. maximum social welfare benchmark]{\label{fig:9b}
\includegraphics[scale=0.29]{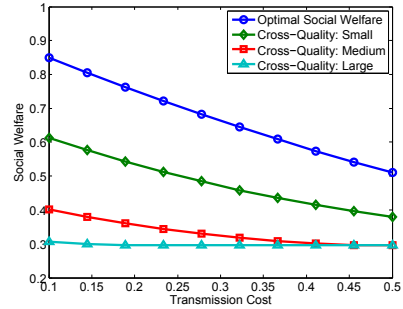}}
~
\subfigure[Social welfare under matching-quality vs. that under cross-quality]{\label{fig:9c}
\includegraphics[scale=0.29]{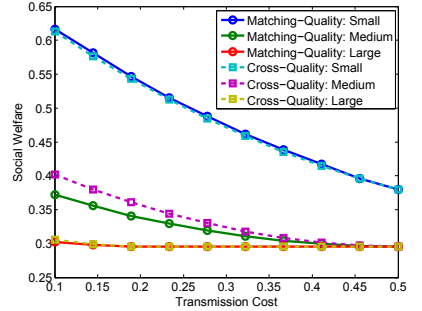}}
\caption{Social welfare comparison under different prices and transmission costs.}\label{WelfareComparison}
\end{figure*}

\subsubsection{Efficiency of Equilibrium}

In this subsection, we compare the social welfare of the equilibrium with the optimal social welfare.
Figs. \ref{Fig:SocialWelfareDifferentParameters}, \ref{Fig:SocialWelfareWholesaleNash}, and \ref{Fig:RelativeWelfare}
show the social welfare of the equilibrium under different system parameters $\eta$, $\wsp$, and $\ctx$. 
We have three observations.

{1).} Figs. \ref{Fig:SocialWelfareDifferentParameters} and \ref{Fig:SocialWelfareWholesaleNash} show that the social welfare decreases in the transmission cost. This is because the generated sharing benefit is reduced when increasing the transmission cost.

{2).} Given the transmission cost $\ctx$, Figs. \ref{Fig:SocialWelfareDifferentParameters} and \ref{Fig:SocialWelfareWholesaleNash} show that the social welfare increases in $\eta$ and decreases in $\wsp$.
This is because a large $\eta$ or a small $\wsp$ will decrease the revenue sharing for the sensing users, thus weakens the competition of the sensing users. In the extreme case $\eta=1$ and $\wsp=0$, the social welfare will achieve the maximum value.

{3).} Fig. \ref{Fig:RelativeWelfare} shows that the ratio of the equilibrium social welfare to the optimal social welfare is increasing in the transmission cost. It demonstrates that a large transmission cost weakens the sharing effect among users.

\subsubsection{Equilibrium Market Shares}
Next, we numerically show the equilibrium market shares.
Figs. \ref{Fig:NashBarMarketSharesSmallRho},  \ref{Fig:NashBarMarketSharesLargeRho}, and  \ref{Fig:WholesaleBarMarketShares} show the results.

{1).} With the increase of the transmission cost, all three figures show that aliens' market share increases and requesters' market share decreases. The reason is that requesters need to pay for the transmission cost. Hence, increasing the transmission cost will change those with high sensing cost to aliens.

{2).} How the sensors' market share changes in cost is more complicated. With a small $\eta$ in Fig.~\ref{Fig:NashBarMarketSharesSmallRho}, the sensor's market share decreases first and then increases to $50\%$; when $\eta$ becomes large enough as in Figs.~\ref{Fig:NashBarMarketSharesLargeRho} and \ref{Fig:WholesaleBarMarketShares}, the sensors' market share monotonically increases to $50\%$. The reason is that, as the transmission cost increases, some requesters will not choose to request data and thus become sensors or aliens. However, when $\eta$ is small as in Fig.~\ref{Fig:NashBarMarketSharesSmallRho}, the decrease of requesters will greatly reduce the sharing benefit of sensors who will obtain $1-\eta$ percentage of the total sharing benefit.
Therefore, some sensors with high sensing cost will also become aliens and thus sensors' market shares decreases, and eventually increases to $50 \%$ in Fig. \ref{Fig:NashBarMarketSharesSmallRho}. The above effect will not play a leading role when $\eta$ is large ($1-\eta$ is small). Therefore, both sensors and aliens monotonically increase as requesters decrease in Figs. \ref{Fig:NashBarMarketSharesLargeRho} and \ref{Fig:WholesaleBarMarketShares}.

{3).} With a very large transmission cost, requesters disappear since there is no point in sharing data. Some autarkic users become sensors and the remaining users become aliens. Thus, sensors and aliens will divide the whole market evenly in Figs. \ref{Fig:NashBarMarketSharesSmallRho}-\ref{Fig:WholesaleBarMarketShares}. The value of $50\%$ is due to our assumption of the joint uniform distribution of value $v$ and cost $c$ in $[0,1]$, which can be verified through (\ref{eq:SW}).

\subsection{Simulations of the Quality-aware Game}
Now we illustrate the efficiency of the quality-aware game equilibrium and the corresponding equilibrium market shares. In comparison with the cross-quality data sharing in Section~\ref{sec:QualitySharing}, we term the quality-aware data sharing in Section~\ref{sec:Quality} as the matching-quality data sharing.

\subsubsection{Quality-aware Maximum Social Welfare Benchmark}
In this subsection, we consider the social welfare maximization problem, where users work together to maximize the sum of all users' payoffs. First, if there is no data sharing, it follows that the social optimality requires those users with $vf(q)\geq cg(q)$ to sense with the quality that
\begin{equation}
q_{k}^{\textsc{s}}(v,c)=\arg\max_{q\in\mathcal{Q}} (vf(q)-cg(q)),
\end{equation}
and those with $vf(q)< cg(q)$ not to sense, due to the requirement of non-negative payoffs. Now we consider incorporating data sharing into the system, with the transmission cost $\ctx$. Those with $vf(q)\geq cg(q)\geq\ctx$ will not sense but acquire data from the sensed users and those with $\ctx<vf(q)<cg(q)$ will also acquire data from the sensed users, due to the payoff $vf(q)-\ctx>vf(q)-cg(q)$. The quality is determined by
\begin{equation}
q_{k}^{\textsc{r}}(v,c)=\arg\max_{q\in\mathcal{Q}} (vf(q)-\ctx).
\end{equation}
Thus, the maximum social welfare is given by
\begin{align}
W_{q}&\textstyle  =\sum\limits_{k=1}^{K}\iint_{\etase_k}(vf(q_{k}^{\textsc{s}}(v,c))-cg(q_{k}^{\textsc{s}}(v,c)))\mathrm{d}v\mathrm{d}c \notag\\
&\textstyle  ~~+\sum\limits_{k=1}^{K}\iint_{\etaex_k}(vf(q_{k}^{\textsc{r}}(v,c))-\ctx)\mathrm{d}v\mathrm{d}c.
\end{align}

\subsubsection{Efficiency of  the Quality-aware Equilibrium}
In this subsection, we compare the equilibrium social welfare with the optimal social welfare benchmark, under the matching-quality and the cross-quality sharing, respectively.
Fig. \ref{WelfareComparison} shows the impacts of the transmission cost and the trading price on the equilibrium social welfare, as well as the equilibrium social welfare comparison results.

First, we can see that the social welfare decreases in the transmission cost in Figs. \ref{fig:9a}-\ref{fig:9c}. This is because the generated sharing benefit is reduced when the transmission cost increases. Second, given the transmission cost, the social welfare  decreases in the  price~$\wsp$  in Figs. \ref{fig:9a}-\ref{fig:9c}. This is because a small price $\wsp$ decreases the revenue sharing for the sensing users, thus weakens the competition among the sensing users. Third, the cross-quality sharing has little impact on the equilibrium social welfare in Fig. \ref{fig:9c}. This is because the cross-quality sharing will not impact requesters' quality selections, which is the determinant part of the social welfare. Furthermore, the cross-quality sharing plays a negative role in the social welfare when the price is small, while it plays a positive role when the price is medium and large in Fig. \ref{fig:9c}. The reason is that the high-quality sensor will first transform the high-quality data into the low-quality data and then transmit to the low-quality data requester with the low-quality price. Hence, the cross-quality data sharing will only be incentivized when the trading price is high, i.e., in the medium and large price scenarios. Furthermore, when the price is sufficiently large, it will discourage the requesters' data requesting decisions. Hence, the social welfare
gap is larger in the medium price scenario, while it is quite small in the large price scenario.

\begin{figure}[h]
\begin{minipage}[t]{0.24\textwidth}
\centering
\includegraphics[scale=0.23]{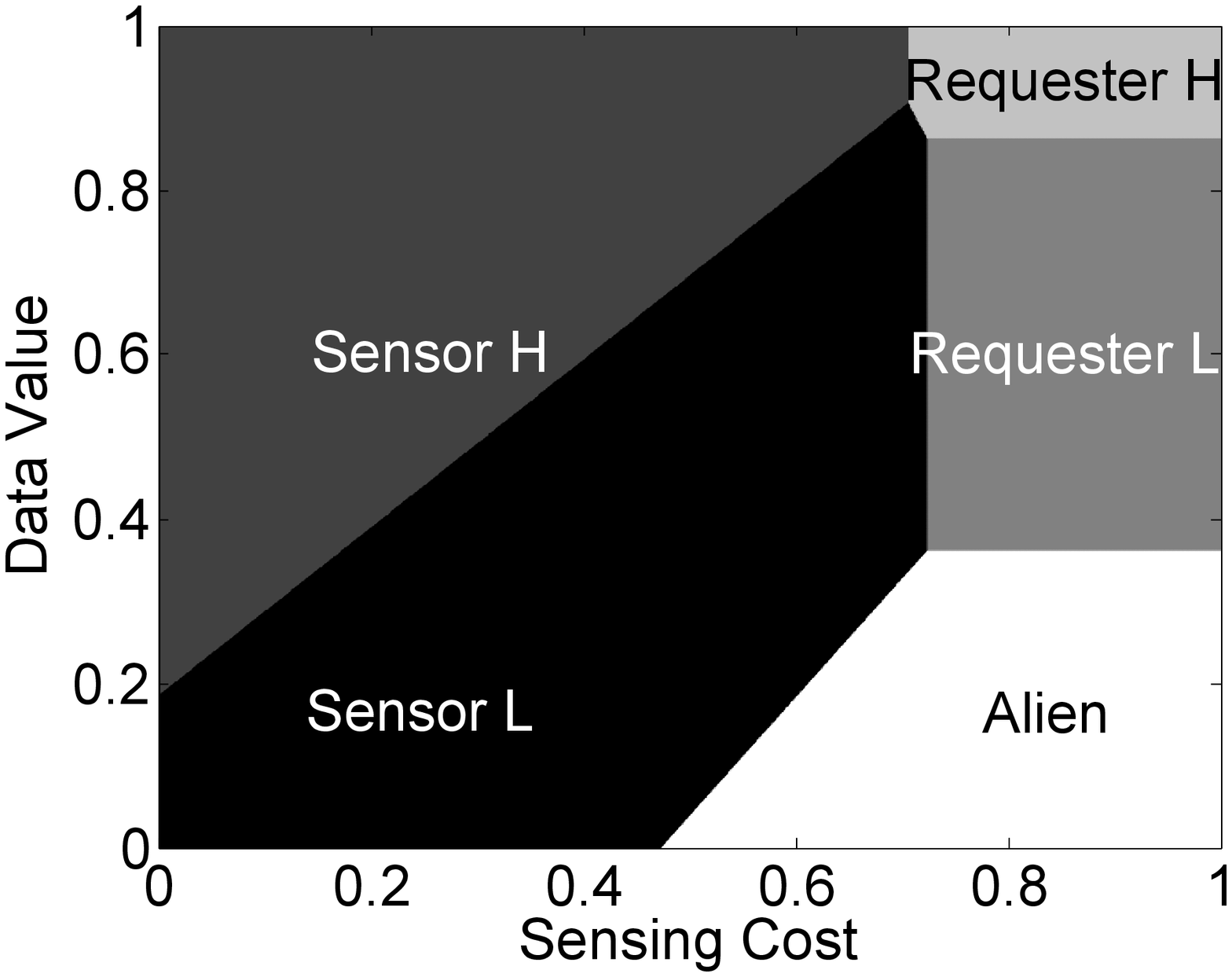}
\caption{Users' selections and partitions: matching-quality sharing.}\label{CPNO}
\end{minipage}
~
\begin{minipage}[t]{0.24\textwidth}
\centering
\includegraphics[scale=0.23]{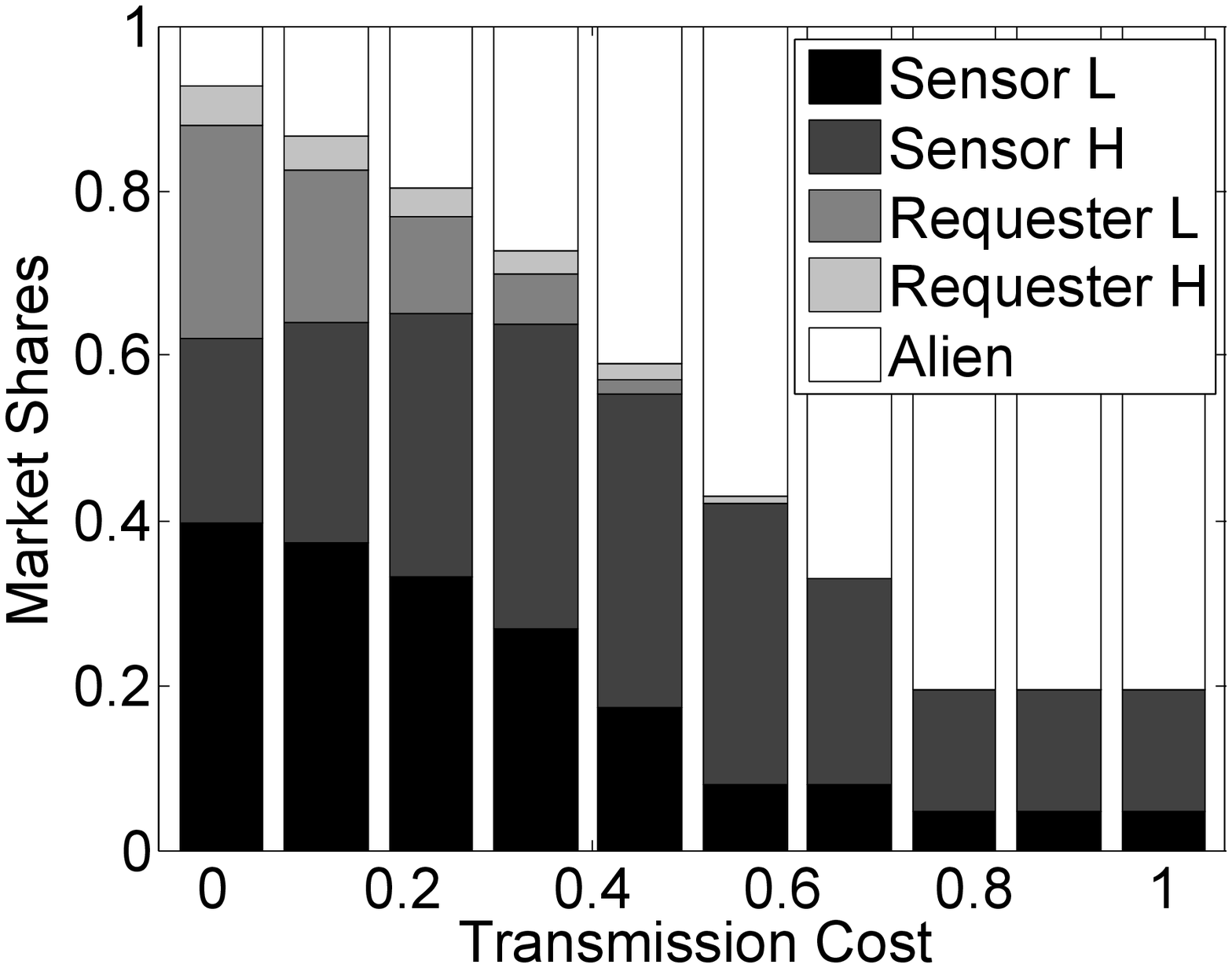}
\caption{Market share vs. transmission cost: matching-quality sharing.}\label{MSNo}
\end{minipage}
\end{figure}

\begin{figure}[h]
\begin{minipage}[t]{0.24\textwidth}
\centering
\includegraphics[scale=0.23]{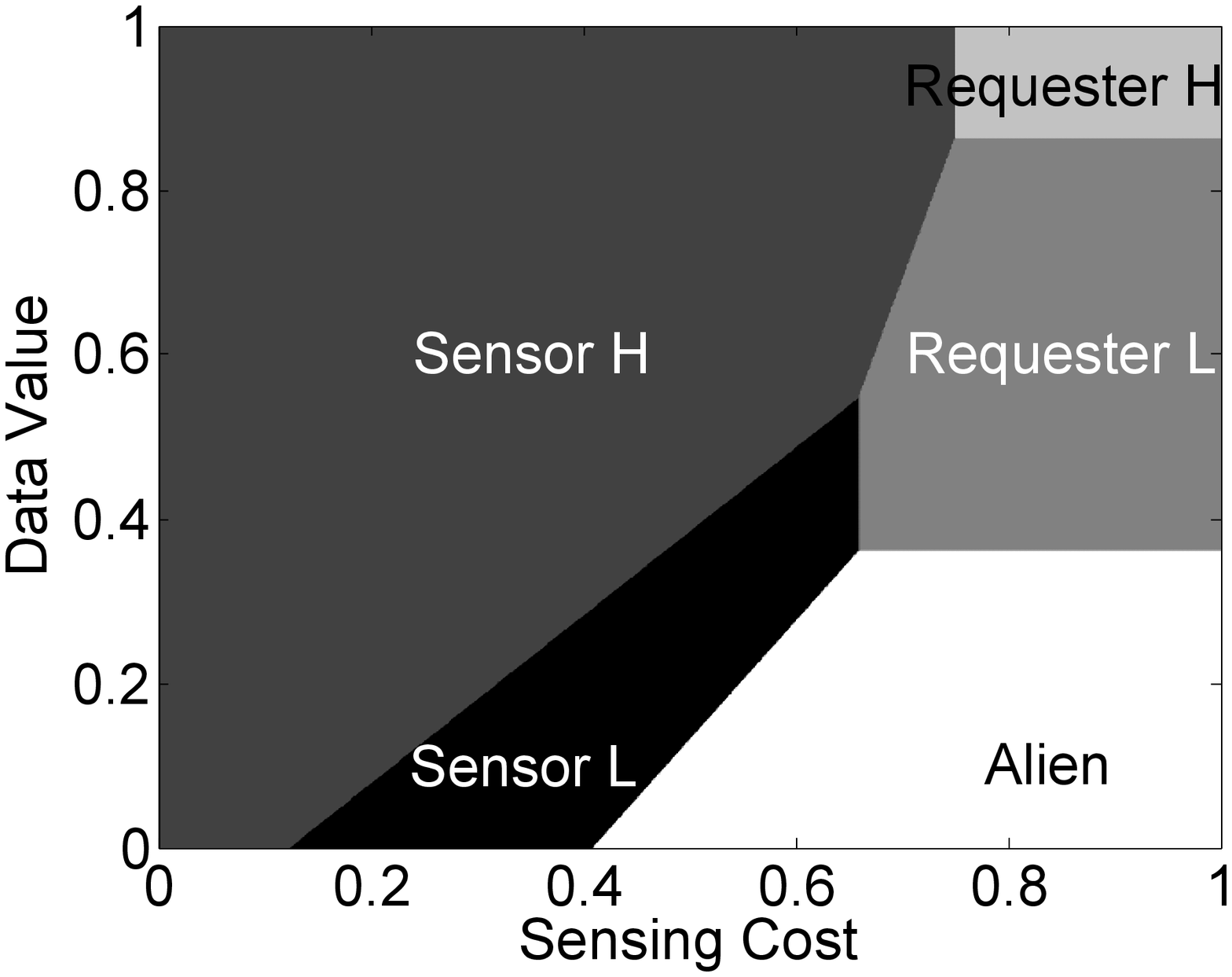}
\caption{Users' selections and partitions: cross-quality sharing.}\label{CPWith}
\end{minipage}
~
\begin{minipage}[t]{0.24\textwidth}
\centering
\includegraphics[scale=0.23]{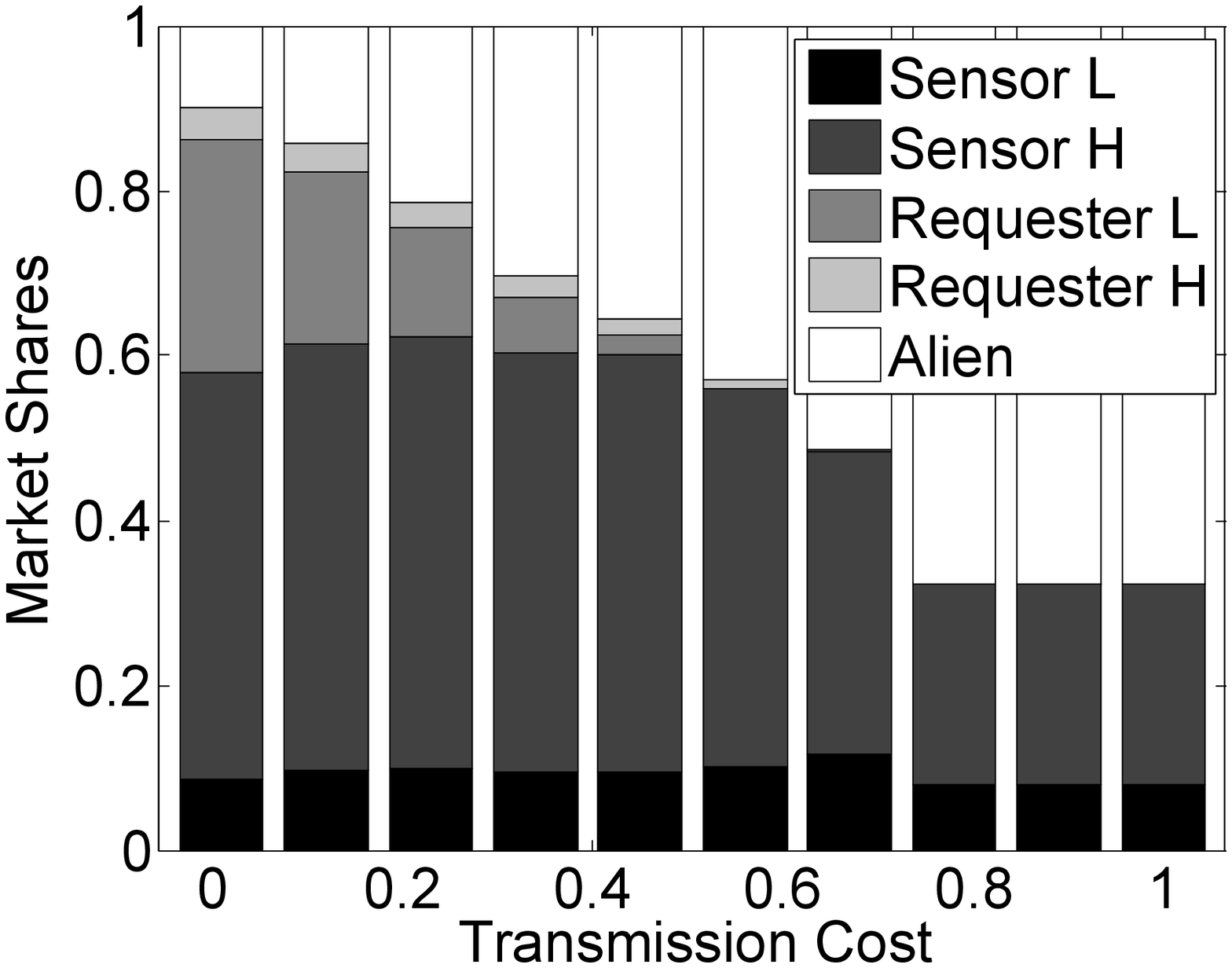}
\caption{Market share vs. transmission cost: cross-quality sharing.}\label{MSWith}
\end{minipage}
\end{figure}

\subsubsection{Quality-aware Equilibrium Market Shares}
In this subsection, we implement Algorithms \ref{algo:max} and \ref{algo:max2} in the quality-aware scenario under the ``matching-quality sharing'' and the ``cross-quality sharing'' settings, respectively.

Figs. \ref{CPNO} and \ref{CPWith} show the users' partitions under the two settings, respectively. 
In Figs. \ref{CPNO} and \ref{CPWith}, ``Sensor L'' denotes sensors obtaining low quality data and  ``Sensor H'' denotes sensors obtaining high quality data, while ``Requester L'' denotes requesters obtaining low quality data and  ``Requester H'' denotes requesters obtaining high quality data.

We have two observations. First, we can see that they follow similar structures as those in Subsection \ref{sec:numerical}, i.e., users' partitions are dependent on the data value and sensing cost distributions. Second, by comparing the dark blue and light blue regions in Figs. \ref{CPNO} and \ref{CPWith}, we can see that the cross-quality sharing will have a great impact on sensors' quality selection, i.e., more sensors choose to sense the high-quality data (Sensor H). However, the cross-quality sharing has little impact on the requesters' quality selection (Requester L and Requester H), due to the fact that a low-quality requesters will receive the requested low quality data.

We further show the the impact of the transmission cost on the equilibrium market shares. Figs. \ref{MSNo} and \ref{MSWith} show the results under matching-quality sharing and cross-quality sharing, respectively, where ``Sensor L'' denotes sensors obtaining low quality data and  ``Sensor H'' denotes sensors obtaining high quality data. From Figs. \ref{MSNo} and \ref{MSWith}, we can see that i) sensors' market share (the summation of Sensor L and Sensor H) first increases with the transmission cost and then decreases with the transmission cost;  ii) requesters' market share  (the summation of Requester L and Requester H) always decreases with the transmission cost; iii) aliens' market share (Alien) always increases with the transmission cost. The non-monotonic change of sensors' market share is due to the interplay of sensors and requesters. When the transmission cost is small and increases, some requesters will become sensors or aliens due to the increase of total cost. Hence, sensors' market share increases. However, when the transmission cost is large and increases, the significant reduction of requesters will reduce the sharing benefit of sensors. Hence, some sensors will become aliens and the market share decreases.

\section{Conclusion}\label{sec:conclusion}
In this work, we present a novel quality-aware P2P-based MCS architecture, which can effectively reduce the management and operational cost on the server.
We study the user behavior dynamics and the market equilibrium of such a system by modeling a data quality-aware non-cooperative game.
In particular, we study the quality-unaware game under a general pricing scheme, and prove the existence and uniqueness of the equilibrium.
Furthermore, we study the quality-aware game under the quality-based pricing scheme, and prove the existence of the equilibrium.
We also propose iterative algorithms that are guaranteed to converge to the game equilibrium under both quality scenarios.
Our numerical results demonstrate the existence and uniqueness of the equilibrium, quantify the efficiency loss, and show the equilibrium market shares under different system parameters and pricing parameters.

There are several possible extensions of the model in this work. For example, it is important to study the coupling of different data when conducting the sensing and sharing decisions. In this case, the interdependence of different data will complicate users' decisions, because each user's utility depends collectively on the correlated data. Moreover, it is also meaningful to consider a dynamic system in a long period of time, considering that the data quality of some applications may decay over time (e.g., the measurement of the air conditions at a time may not be accurate at later times). Hence, it is challenging to analyze users' decisions due to the time domain coupling.

\appendices
\section{Proof of Proposition 1}
\begin{IEEEproof}
Notice that there is a one-to-one mapping between a strategy profile $\{ x(v,c), \forall v,c \}$ and an average sharing benefit $\delta$. The reason is that, given any strategy profile $\{ x(v,c), \forall v,c \}$, the market shares $\widetilde{\etase}$, $\widetilde{\etaex}$, and $ \widetilde{\etanl}$ will be uniquely determined. Then we can determine the new average sharing benefit $\widetilde{\delta}$ according to Eq. (17). Therefore, if the strategy profile $\{ x^{\ast}(v,c) , \forall v,c \}$ is a stable point, it follows that the average sharing benefit $\delta =  \frac{B^{\textsc{se}}(\delta)}{N^{\textsc{se}}(\delta)}$. Otherwise, the average sharing benefit $\delta$ will deviate from  $\frac{B^{\textsc{se}}(\delta)}{N^{\textsc{se}}(\delta)}$, and the corresponding strategy profile $\{ x(v,c) , \forall v,c \}$ will also change. This completes the proof.
\end{IEEEproof}

\section{Proof of Proposition 2}
\textbf{Case 1. High Average Sharing Benefit.}  In this case $\delta\geq \delta_0$. We calculate $N^{\textsc{se}}_{h}(\delta)$ and $B^{\textsc{se}}_{h}(\delta)$ by integrating the two integrals $B^{\textsc{se}}(\delta)$ and $N^{\textsc{se}}(\delta)$, respectively. That is,
$$
\begin{aligned}
& B^{\textsc{se}}_{h}(\delta)\\
&\textstyle=N\left[\frac{1}{6w(1-\rho)}\left(1-\ctx-\frac{\wsp}{\rho}-\delta\right)^3+\wsp\frac{(1-\ctx-\frac{\wsp}{\rho}-\delta)(1-\delta-\rho\ctx-\wsp)}{2w(1-\rho)}\right].
\end{aligned}
$$
$$
\begin{aligned}
N^{\textsc{se}}_{h}(\delta)&\textstyle=N\left[1-\frac{(1-\delta-\rho\ctx-\wsp)^2}{2w(1-\rho)}-\frac{\rho(\ctx+\frac{\wsp}{\rho})^2}{2w}\right].
\end{aligned}
$$
First notice that $w$ is just a scalar in Fig. 2. Hence, the analysis results will not change if we normalize $w$ as 1. The parameter $N$ will not impact the equilibrium, since $N$ is canceled according to Proposition 1. We will omit $N$ when deriving the equilibrium $\delta$. Hence, the function $\Lambda(\delta)$ is given by
$$
\begin{aligned}
\Lambda(\delta)=\Lambda_h(\delta)&=\delta N^{\textsc{se}}_{h}(\delta)-B^{\textsc{se}}_{h}(\delta)\\
&=\delta\left[1-\frac{(1-\delta-\rho\ctx-\wsp)^2}{2(1-\rho)}-\frac{\rho(\ctx+\frac{\wsp}{\rho})^2}{2}\right]\\
&~~~~~~-\frac{1}{6(1-\rho)}(1-\ctx-\frac{\wsp}{\rho}-\delta)^3\\
&~~~~~~-\wsp\frac{(1-\ctx-\frac{\wsp}{\rho}-\delta)(1-\delta-\rho\ctx-\wsp)}{2(1-\rho)}.
\end{aligned}
$$

\begin{lemma}
The function $\Lambda_h(\delta)$ is monotonically increasing in $\delta$ when $\delta\geq\delta_0$.
\end{lemma}
\begin{IEEEproof}
We derive the first-order derivative of $\Lambda_h(\delta)$ with respect to $\delta$, i.e.,
$$
\begin{aligned}
\Lambda_h'(\delta)&=\left[1-\frac{(1-\delta-\rho\ctx-\wsp)^2}{2(1-\rho)}-\frac{\rho(\ctx+\frac{\wsp}{\rho})^2}{2}\right]\\
&~~~~~~+\delta\frac{2(1-\delta-\rho\ctx-\wsp)}{2(1-\rho)}+\frac{3(1-\delta-\ctx-\frac{\wsp}{\rho})^2}{6(1-\rho)}\\
&~~~~~~+\frac{\wsp}{2(1-\rho)}\left(2-2\delta-(1+\rho)(\ctx+\frac{\wsp}{\rho})\right)\\
&=1-\frac{\rho(\ctx+\frac{\wsp}{\rho})^2}{2}+\delta\frac{1-\delta-\rho\ctx-\wsp}{1-\rho}\\
&~~~~~~~-\frac{(\ctx+\frac{\wsp}{\rho})(2-2\delta-(1+\rho)(\ctx+\frac{\wsp}{\rho}))}{2}\\
&~~~~~~~+\frac{\wsp}{2(1-\rho)}\left(2-2\delta-(1+\rho)(\ctx+\frac{\wsp}{\rho})\right).\\
\end{aligned}
$$

First, we can see that $\Lambda_h'(\delta)$ is a parabola going downwards with $$\Lambda_h'(0)=\frac{(1-\ctx-\frac{\wsp}{\rho})^2}{2}+\frac{\wsp}{2(1-\rho)}\left(2-(1+\rho)(\ctx+\frac{\wsp}{\rho})\right)\geq0.$$
The inequality follows from $\frac{\wsp}{2(1-\rho)}\left(2-(1+\rho)(\ctx+\frac{\wsp}{\rho})\right)\geq\frac{\wsp}{2(1-\rho)}\left(2(1-\ctx-\frac{\wsp}{\rho})\right)\geq0$, since $1+\rho\leq2$ and $1-\ctx-\frac{\wsp}{\rho}\geq0$.
Moreover, we have
$$
\begin{aligned}
&\Lambda_h'(1-\ctx-\frac{\wsp}{\rho})\\
&=\left(\ctx+\frac{\wsp}{\rho}\right)\left(1-\ctx-\frac{\wsp}{\rho}\right)+\frac{2-(\ctx+\frac{\wsp}{\rho})^2}{2}+\frac{\wsp}{2}\left(\ctx+\frac{\wsp}{\rho}\right)\geq0.
\end{aligned}
$$ Hence, we have $\Lambda_h'(\delta)>0$ for all $\delta\geq\delta_0$. This completes the proof.
\end{IEEEproof}

Based on the above, we now prove Proposition 2.
\begin{IEEEproof}
Fist, we claim that $\Lambda_h(1-\ctx-\frac{\wsp}{\rho})>0$. To see this, we substitute $1-\ctx-\frac{\wsp}{\rho}$ into $\Lambda_h(\delta)$, we have $\Lambda_h(1-\ctx-\frac{\wsp}{\rho})=(1-\ctx-\frac{\wsp}{\rho})(1-\frac{(\ctx+\frac{\wsp}{\rho})^2}{2})>0$. Second, according to Sec. 2.1, the function $\Lambda_h(\delta)$ is monotonically increasing, hence, the equation $\Lambda_h(\delta)=0$ has a unique solution on $(\delta_0,1-\ctx-\frac{\wsp}{\rho})$ if and only if $\Lambda_h(\delta_0)<0$, and has no solution if and only if $\Lambda_h(\delta_0)\geq0$. By substituting $\delta_0$ into $\Lambda_h(\delta)$, we have
$$
\begin{aligned}
&\Lambda_h(\delta_0)\\
&=\frac{(1-s-\frac{\wsp}{\rho})3\rho(1+\rho-\rho(\ctx+\frac{\wsp}{\rho})^2)}{6}\\
&~~~~~~~-\frac{(1-\ctx-\frac{\wsp}{\rho})^3(1-\rho)^2}{6}\\
&~~~~~~~-\frac{\wsp}{2}(1-\rho)(1-\ctx-\frac{\wsp}{\rho}).
\end{aligned}
$$
Thus, $\Lambda_h(\delta_0)<0$ if and only if $3\rho(1+\rho-\rho(\ctx+\frac{\wsp}{\rho})^2)<(1-\ctx-\frac{\wsp}{\rho})^2(1-\rho)^2+3\wsp(1-\rho)$ and $\Lambda_h(\delta_0)\geq0$ if and only if $3\rho(1+\rho-\rho(\ctx+\frac{\wsp}{\rho})^2)\geq(1-\ctx-\frac{\wsp}{\rho})^2(1-\rho)^2+3\wsp(1-\rho)$. This completes the proof.
\end{IEEEproof}
\section{Proof of Proposition 3}
\textbf{Case 2. Low Average Sharing Benefit.} In this case $0\leq \delta\leq \delta_0$. We calculate $B^{\textsc{se}}(\delta)$ and $N^{\textsc{se}}(\delta)$ as $B^{\textsc{se}}_l(\delta)$ and $N^{\textsc{se}}_l(\delta)$ similarly. That is,
$$
\begin{aligned}
& B^{\textsc{se}}_l(\delta)\\
&=N\iint_{\etaex(v,c)} (1-\rho)(w\cdot v\cdot -\ctx)f_{vc}(v,c)\mathrm{d}v\mathrm{d}c\\
&~~~~~~~~~~+N\wsp\iint_{\etaex(v,c)}f_{vc}(v,c)\mathrm{d}v\mathrm{d}c\\
&=N\left[\frac{(1-\rho)^2(w-\ctx-\frac{\wsp}{\rho})^3}{6w}\right.\\
&\left.+(1-\rho)\frac{(w-\ctx-\frac{\wsp}{\rho})^2}{2w}(1-w(1-\rho)-\delta-\rho \ctx-\wsp)\right]\\
&+N\wsp\frac{w-\ctx-\frac{\wsp}{\rho}}{2w}\left(1-w(1-\rho)-\delta-\rho \ctx-\wsp+1-2\ctx-\frac{\wsp}{\rho}\right).
\end{aligned}
$$
$$
\begin{aligned}
N^{\textsc{se}}_l(\delta) 
&=N\left[\delta+\rho \ctx+\wsp+\frac{w(1-\rho)}{2}-\frac{\rho(\ctx+\frac{\wsp}{\rho})^2}{2w}\right].
\end{aligned}
$$

Similarly, we normalize $w$ as 1 and omit $N$.
Hence, the function $\Lambda_l(\delta)$ is given by
$$
\begin{aligned}
&\Lambda_l(\delta)\\
&\textstyle=\delta\left[\delta+\rho \ctx+\wsp+\frac{(1-\rho)}{2}-\frac{\rho(\ctx+\frac{\wsp}{\rho})^2}{2}\right]-\frac{(1-\rho)^2(1-\ctx-\frac{\wsp}{\rho})^3}{6}\\
&~~~~~~-(1-\rho)\frac{(1-\ctx-\frac{\wsp}{\rho})^2}{2}(1-(1-\rho)-\delta-\rho \ctx-\wsp)\\
&~~~~~~-\wsp\frac{1-\ctx-\frac{\wsp}{\rho}}{2}\left(1-(1-\rho)-\delta-\rho \ctx-\wsp+1-2\ctx-\frac{\wsp}{\rho}\right)\\
&\textstyle=(\delta)^2+\left((1+\frac{\wsp}{2})(1-\ctx-\frac{\wsp}{\rho})+\frac{(\ctx+\frac{\wsp}{\rho})^2}{2}-\rho(1-\ctx-\frac{\wsp}{\rho})^2\right)\delta\\
&~~~~~~-\frac{1}{6}(1-\rho)(1+2\rho)(1-\ctx-\frac{\wsp}{\rho})^3\\
&~~~~~~-\frac{\wsp(1-\ctx-\frac{\wsp}{\rho})}{2}\left((1+\rho)(1-\ctx-\frac{\wsp}{\rho})-\ctx\right).
\end{aligned}
$$
\begin{lemma}
The function $\Lambda_l(\delta)$ is either monotonically increasing with $\delta\in[0,\delta_0]$, or first decreasing with $\delta$ and then increasing with $\delta\in[0,\delta_0]$.
\end{lemma}
\begin{IEEEproof}
We first check the first-order derivative of $\Lambda_l(\delta)$ with respect to $\delta$, i.e.,
$$
\Lambda_l'(\delta)=2\delta+(1+\frac{\wsp}{2})(1-\ctx-\frac{\wsp}{\rho})+\frac{(\ctx+\frac{\wsp}{\rho})^2}{2}-\rho(1-\ctx-\frac{\wsp}{\rho})^2.
$$
When $\delta=\delta_0$, we have
$$
\begin{aligned}
\Lambda_l'(\delta_0)&=2\rho(1-\ctx-\frac{\wsp}{\rho})+(1+\frac{\wsp}{2})(1-\ctx-\frac{\wsp}{\rho})+\frac{(\ctx+\frac{\wsp}{\rho})^2}{2}\\
&~~~~~~~-\rho(1-\ctx-\frac{\wsp}{\rho})^2\\
&=\rho(1-\ctx-\frac{\wsp}{\rho})(1+\ctx+\frac{\wsp}{\rho})+\frac{1+(1-\ctx-\frac{\wsp}{\rho})^2}{2}\\
&~~~~~~~+\frac{\wsp}{2}(1-\ctx-\frac{\wsp}{\rho})>0.
\end{aligned}
$$
When $\delta=0$, we have
$$
\begin{aligned}
\Lambda_l'(0)&=(1+\frac{\wsp}{2})(1-\ctx-\frac{\wsp}{\rho})+\frac{(\ctx+\frac{\wsp}{\rho})^2}{2}-\rho(1-\ctx-\frac{\wsp}{\rho})^2\\
&=(\frac{1}{2}-\rho)(1-\ctx-\frac{\wsp}{\rho})^2+\frac{1}{2}+\frac{\wsp}{2}(1-\ctx-\frac{\wsp}{\rho}).
\end{aligned}
$$
Hence, $\Lambda_l'(0)\geq0$ if and only if $(\frac{1}{2}-\rho)(1-\ctx-\frac{\wsp}{\rho})^2+\frac{1}{2}+\frac{\wsp}{2}(1-\ctx-\frac{\wsp}{\rho})\geq0$, which completes the first part.
When $(\frac{1}{2}-\rho)(1-\ctx-\frac{\wsp}{\rho})^2+\frac{1}{2}+\frac{\wsp}{2}(1-\ctx-\frac{\wsp}{\rho})<0$, the unique solution to $\Lambda_l'(\delta)=0$ is $-\frac{(\frac{1}{2}-\rho)(1-\ctx-\frac{\wsp}{\rho})^2+\frac{1}{2}+\frac{\wsp}{2}(1-\ctx-\frac{\wsp}{\rho})}{2}$, which shows that $\Lambda_l'(\delta)\leq0$ on $[0,-\frac{(\frac{1}{2}-\rho)(1-\ctx-\frac{\wsp}{\rho})^2+\frac{1}{2}+\frac{\wsp}{2}(1-\ctx-\frac{\wsp}{\rho})}{2}]$ and $\Lambda_l'(\delta)\geq0$ on $[-\frac{(\frac{1}{2}-\rho)(1-\ctx-\frac{\wsp}{\rho})^2+\frac{1}{2}+\frac{\wsp}{2}(1-\ctx-\frac{\wsp}{\rho})}{2},\delta_0]$. This completes the proof.
\end{IEEEproof}

Based on the above, we now prove Proposition 2.
\begin{IEEEproof}
When $\delta=0$, we have
$$
\begin{aligned}
\Lambda_l(0)&=-\frac{1}{6}(1-\rho)(1+2\rho)(1-\ctx-\frac{\wsp}{\rho})^3\\
&~~~~~~~-\frac{\wsp(1-\ctx-\frac{\wsp}{\rho})}{2}\left((1+\rho)(1-\ctx-\frac{\wsp}{\rho})-\ctx\right)\leq0.
\end{aligned}
$$
Then there exist at least one equilibrium $\delta^{\ast}$ such that $\Lambda_l(\delta^{\ast})=0$ if and only if $\Lambda_l(\delta_0)>0$, which yields
$$
\begin{aligned}
\Lambda_l(\delta_0)&=\frac{1-\ctx-\frac{\wsp}{\rho}}{6}\left[2(1-\ctx-\frac{\wsp}{\rho})(1+2(\ctx+\frac{\wsp}{\rho}))\rho^2\right.\\
&~~~~~~~\left.+(2(1-\ctx-\frac{\wsp}{\rho})^2+3)\rho-(1-\ctx-\frac{\wsp}{\rho})^2\right]\\
&~~~~~~~-\frac{\wsp(1-\ctx-\frac{\wsp}{\rho})}{2}\left((1+\rho)(1-\ctx-\frac{\wsp}{\rho})-\ctx\right)>0.
\end{aligned}
$$

Furthermore, the condition $\Lambda_l(\delta_0)>0$ will guarantee that the equilibrium solution is unique. The reason is that the function $\Lambda_l(\delta)$ is either monotonically increasing or monotonically decreasing first then monotonically increasing, taking into account $\Lambda_l(0)\leq0$ and $\Lambda_l(\delta_0)>0$. This completes the proof.
\end{IEEEproof}

\section{Proof of Theorem 1}
\begin{IEEEproof}
By combining the results in Propositions 2 and 3, we conclude that the existence can be readily guaranteed, which is either in $[0,\delta_0]$ or in $[\delta_0,+\infty)$. The corresponding conditions that guarantee the existence and uniqueness are \emph{mutually exclusive and exhaustive} to each other in Propositions 2 and 3, which shows that Theorem \ref{theoremgeneral} immediately holds. This completes the proof.
\end{IEEEproof}

\section{Proof of Proposition 4}
\begin{IEEEproof}
We consider the Banach Fixed Point Theorem for contraction mapping. The function $f(\delta)$ is defined as
$$
f(\delta)=\left\{
\begin{aligned}
& \lambda\delta+(1-\lambda)\frac{B^{\textsc{se}}_l(\delta)}{N^{\textsc{se}}_l(\delta)}, \text{~~ if ~~} 0\leq \delta\leq \delta_0,\\
& \lambda\delta+(1-\lambda)\frac{B^{\textsc{se}}_h(\delta)}{N^{\textsc{se}}_h(\delta)}, \text{~~ if ~~} \delta\geq\delta_0.
\end{aligned}
\right.
$$
We next show that the function $f(\delta)$ is a contraction mapping.

\textbf{1).} We first consider the interval $[0,\delta_0]$, where the function $f$ is given by
$$
\begin{aligned}
f(\delta)&=\lambda\delta\\
&\textstyle+(1-\lambda)\frac{\frac{(1-\rho)^2(1-\ctx-\frac{\wsp}{\rho})^3}{6}+(1-\rho)\frac{(1-\ctx-\frac{\wsp}{\rho})^2}{2}(1-(1-\rho)-\delta-\rho \ctx-\wsp)}{\delta+\rho \ctx+\wsp+\frac{(1-\rho)}{2}-\frac{\rho(\ctx+\frac{\wsp}{\rho})^2}{2}}\\
&\textstyle+(1-\lambda)\frac{\wsp\frac{1-\ctx-\frac{\wsp}{\rho}}{2}\left(1-(1-\rho)-\delta-\rho \ctx-\wsp+1-2\ctx-\frac{\wsp}{\rho}\right)}{\delta+\rho \ctx+\wsp+\frac{(1-\rho)}{2}-\frac{\rho(\ctx+\frac{\wsp}{\rho})^2}{2}}.
\end{aligned}
$$
The derivative with respect to $\delta$ is
$$
\begin{aligned}
f'(\delta)&\textstyle=\lambda+(1-\lambda)\frac{-\frac{(1-\rho)(1-\ctx-\frac{\wsp}{\rho})^2}{2}\left(\frac{1+\rho-\rho(\ctx+\frac{\wsp}{\rho})^2}{2}+\frac{(1-\rho)(1-\ctx-\frac{\wsp}{\rho})}{3}\right)}{\left(\delta+\rho \ctx+\wsp+\frac{(1-\rho)}{2}-\frac{\rho(\ctx+\frac{\wsp}{\rho})^2}{2}\right)^2}\\
&\textstyle+(1-\lambda)\frac{-\wsp\frac{1-\ctx-\frac{\wsp}{\rho}}{2}\left(2-\frac{(1-\rho)}{2}-\frac{\rho(\ctx+\frac{\wsp}{\rho})^2}{2}-2\ctx-\frac{\wsp}{\rho}\right)}{\left(\delta+\rho \ctx+\wsp+\frac{(1-\rho)}{2}-\frac{\rho(\ctx+\frac{\wsp}{\rho})^2}{2}\right)^2}.
\end{aligned}
$$
Since the function $f'(\delta)$ is an increasing function of $\delta$, we have $f'(0)\leq f'(\delta)\leq f'(\delta_0)$, moreover, we have
$$
| f'(\delta)|\leq\max\left(|f'(0)|,|f'(\delta_0)|\right).
$$
Plugging $0$ and $\delta_0$ into $f'(\delta)$, we have
$$
|f'(0)|=\left|\lambda+(1-\lambda)\frac{-P_1}{P_2}\right|,
$$
and
$$
|f'(\delta_0)|=\left|\lambda+(1-\lambda)\frac{-P_1}{P_3}\right|,
$$
where we define $$
\begin{aligned} P_1&\textstyle=\frac{(1-\rho)(1-\ctx-\frac{\wsp}{\rho})^2}{2}\left(\frac{1+\rho-\rho(\ctx+\frac{\wsp}{\rho})^2}{2}+\frac{(1-\rho)(1-\ctx-\frac{\wsp}{\rho})}{3}\right)\\
&\textstyle+\wsp\frac{1-\ctx-\frac{\wsp}{\rho}}{2}\left(\frac{1+\rho-\rho(\ctx+\frac{\wsp}{\rho})^2}{2}+1-2\ctx-\frac{\wsp}{\rho}\right),
\end{aligned}
$$
$$P_2=\left(\rho \ctx+\wsp+\frac{(1-\rho)}{2}-\frac{\rho(\ctx+\frac{\wsp}{\rho})^2}{2}\right)^2,$$ and $$P_3=\left(\frac{1+\rho-\rho(\ctx+\frac{\wsp}{\rho})^2}{2}\right)^2.$$ Since $0\leq\rho \ctx+\frac{(1-\rho)}{2}-\frac{\rho(\ctx+\frac{\wsp}{\rho})^2}{2}\leq \frac{1+\rho-\rho(\ctx+\frac{\wsp}{\rho})^2}{2}$, we have $P_2\leq P_3$. Hence, $\frac{P_1}{P_2}\geq\frac{P_1}{P_3}$.

Hence, $| f'(\delta)|<1\Leftrightarrow\max(|f'(0)|,|f'(\delta_0)|)<1$, which means that
$$
\begin{aligned}
-1&<\lambda-(1-\lambda)\frac{P_1}{P_2}&<1,\\
-1&<\lambda-(1-\lambda)\frac{P_1}{P_3}&<1.
\end{aligned}
$$
We have
$$
\frac{1+\lambda}{1-\lambda}>\frac{P_1}{P_2}.
$$
Hence, if $P_1<P_2$, we can just set $\lambda=0$; if $P_1\geq P_2$, we can set $\lambda=\frac{P_1-P_2}{P_1+P_2}+\epsilon$ for a sufficiently small positive $\epsilon$.

For any $\delta_1,\delta_2\in [0,\delta_0]$, where without loss of generality $\delta_1<\delta_2$, we have by the mean value theorem that there exists $\xi \in (\delta_1,\delta_2)$ such that
$f(\delta_1)-f(\delta_2) = f'(\xi)(\delta_1-\delta_2) \Rightarrow |f(\delta_1)-f(\delta_2)| = |f'(\xi)||(\delta_1-\delta_2)|$. This shows that the function $f$ is a contraction mapping with factor bounded by $|f'(\xi)|$ if we choose $\lambda$ according to the above rule, i.e., $\lambda=0$ for $P_1<P_2$ and $\lambda=\frac{P_1-P_2}{P_1+P_2}+\epsilon$ for $P_1\geq P_2$.

Hence, in this case the threshold $\lambda_0$ is given by
$$
\lambda_0 = \max\left( \frac{-\psi_l(0)-1}{-\psi_l(0)+1}, \  0\right),
$$
where $\psi_l(0)=-P_1/P_2<1$.

\textbf{2).} Now we consider the other interval, i.e., $[\delta_0,1-\ctx-\frac{\wsp}{\rho}]$, where the function $f$ is given by
$$
\textstyle f(\delta)=\lambda\delta+(1-\lambda)\frac{(1-\ctx-\frac{\wsp}{\rho}-\delta)^3+3\wsp(1-\ctx-\frac{\wsp}{\rho}-\delta)(1-\delta-\rho(\ctx+\frac{\wsp}{\rho}))}{3(1-\rho)(2-\rho(\ctx+\frac{\wsp}{\rho})^2)-3(1-\delta-\rho(\ctx+\frac{\wsp}{\rho}))^2}.
$$
The derivative with respect to $\delta$ is
$$
\begin{aligned}
&\textstyle f'(\delta)=\lambda+(1-\lambda)\frac{-3(1-\ctx-\frac{\wsp}{\rho}-\delta)^2}{\left[3(1-\rho)(2-\rho\ctx^2)-3(1-\delta-\rho\ctx)^2\right]^2}\\
&\textstyle~~~~~~~~~~~~~~~~\times\left[3(1-\rho)(2-\rho(\ctx+\frac{\wsp}{\rho})^2)\right.\\
&\textstyle~~~~~~~~~~~~~~~~~~\left.-(1-\delta-\rho\ctx-\wsp)(1-\delta+(2-3\rho)(\ctx+\frac{\wsp}{\rho}))\right]\\
&\textstyle~~+(1-\lambda)\frac{9\wsp(1-\rho)\left[(2-\rho\ctx^2)(2\delta+(1+\rho)(\ctx)-2)+\ctx((1-\delta-\rho\ctx)^2)\right]}{\left[3(1-\rho)(2-\rho\ctx^2)-3(1-\delta-\rho\ctx)^2\right]^2}.
\end{aligned}
$$
Since the function $f'(\delta)$ is an increasing function of $\delta$, we have $f'(\delta_0)\leq f'(\delta)\leq f'(1-\ctx-\frac{\wsp}{\rho})$, moreover, we have
$$
| f'(\delta)|\leq\max\left(|f'(\delta_0)|,|f'(1-\ctx-\frac{\wsp}{\rho})|\right).
$$
Plugging $\delta_0$ and $1-\ctx-\frac{\wsp}{\rho}$ into $f'(\delta)$, we have
$$
\begin{aligned}
|f'(\delta_0)|&\textstyle=\left|\lambda+(1-\lambda)\frac{1}{3\left(1+\rho-\rho\ctx^2\right)^2}\right.\\
&~~~~~~\textstyle\left.\times\left[-(1-\ctx)^2(1-\rho)(5-2\ctx+\rho(1-\ctx)(1+3\ctx))\right.\right.\\
&~~~~~~~~~~~~~~~~~~~~~~~~\textstyle\left.\left.-3\wsp[(2-\rho\ctx^2)(2-\ctx)-(1-\rho)\ctx]\right.\right|.
\end{aligned}
$$
$$
|f'(1-\ctx-\frac{\wsp}{\rho})|=\left|\lambda+(1-\lambda)\frac{-\wsp(\ctx+\frac{\wsp}{\rho})}{2-(\ctx+\frac{\wsp}{\rho})^2}\right|.
$$
We define
$$
\begin{aligned}
Q_1&=(1-\ctx-\frac{\wsp}{\rho})^2(1-\rho)\\
&~~~~~~~~~~\times[5-2(\ctx+\frac{\wsp}{\rho})+\rho(1-\ctx-\frac{\wsp}{\rho})(1+3\ctx+\frac{3\wsp}{\rho})]\\
&~~~+3\wsp[(2-\rho(\ctx+\frac{\wsp}{\rho})^2)(2-\ctx-\frac{\wsp}{\rho})-(1-\rho)(\ctx+\frac{\wsp}{\rho})],
\end{aligned}
$$
and
$$Q_2=3\left(1+\rho-\rho(\ctx+\frac{\wsp}{\rho})^2\right)^2.$$

Hence, $| f'(\delta)|<1\Leftrightarrow\max(|f'(\delta_0)|,|f'(1-\ctx-\frac{\wsp}{\rho})|)<1$, which means that
$$
\begin{aligned}
&-1<\lambda-(1-\lambda)\frac{Q_1}{Q_2}<1,\\
&-1<\lambda-(1-\lambda)\frac{\wsp(\ctx+\frac{\wsp}{\rho})}{2-(\ctx+\frac{\wsp}{\rho})^2}<1.
\end{aligned}
$$
We have
$$
\frac{1+\lambda}{1-\lambda}>\frac{Q_1}{Q_2} \text{ and } \frac{1+\lambda}{1-\lambda}>\frac{\wsp(\ctx+\frac{\wsp}{\rho})}{2-(\ctx+\frac{\wsp}{\rho})^2}.
$$
Note that the latter inequality always holds due to $\frac{\wsp(\ctx+\frac{\wsp}{\rho})}{2-(\ctx+\frac{\wsp}{\rho})^2}\leq1$. Hence, if $Q_1<Q_2$, we can just set $\lambda=0$; if $Q_1\geq Q_2$, we can set $\lambda=\frac{Q_1-Q_2}{Q_1+Q_2}+\epsilon$ for a sufficiently small positive $\epsilon$.

For any $\delta_1,\delta_2\in [\delta_0,1-\ctx-\frac{\wsp}{\rho}]$, where without loss of generality $\delta_1<\delta_2$, we have by the mean value theorem that there exists $\xi \in (\delta_1,\delta_2)$ such that
$f(\delta_1)-f(\delta_2) = f'(\xi)(\delta_1-\delta_2) \Rightarrow |f(\delta_1)-f(\delta_2)| = |f'(\xi)||(\delta_1-\delta_2)|$. This shows that the function $f$ is a contraction mapping with factor bounded by $|f'(\xi)|$ if we choose $\lambda$ according to the above rule, i.e., $\lambda=0$ for $Q_1<Q_2$ and $\lambda=\frac{Q_1-Q_2}{Q_1+Q_2}+\epsilon$ for $Q_1\geq Q_2$.
Hence, in this case the threshold $\lambda_0$ is given by
$$
\lambda_0 = \max\left( \frac{-\psi_h(\delta_0)-1}{-\psi_h(\delta_0)+1},  \ 0\right),
$$
where $\psi_h(\delta_0)=-Q_1/Q_2<1$.

\textbf{3).} According to Banach fixed point theorem for contraction mapping, there exists a unique $\delta$ such that $f(\delta)=\delta$, which completes the convergence proof.
\end{IEEEproof}

\section{Proof of Proposition 5}

\begin{IEEEproof}
Notice that there is a one-to-one mapping between a strategy profile $\{(x(v,c), q(v,c)),\forall v,c \}$ and an average sharing benefit $\delta_k$. The reason is that, given any strategy profile $\{ (x(v,c), q(v,c)), \forall v,c \}$, the market shares $\widetilde{\etase_k}$, $\widetilde{\etaex_k}$, and $ \widetilde{\etanl}$ will be uniquely determined. Then we can determine the new average sharing benefit $\widetilde{\delta_k}$ according to Eq. (36). Therefore, if the strategy profile $\{ ( x^{\ast}(v,c),q^{\ast}(v,c) ), \forall v,c \}$ is a stable point, it follows that the average sharing benefit $\delta_k =   \wsp\cdot h(q_k)\cdot \frac{|\widetilde{\etaex_k}(\boldsymbol{\Phi})|}{|\widetilde{\etase_k}(\boldsymbol{\Phi})|}$. Otherwise, the average sharing benefit $\delta_k$ will deviate from  $ \wsp\cdot h(q_k)\cdot \frac{|\widetilde{\etaex_k}(\boldsymbol{\Phi})|}{|\widetilde{\etase_k}(\boldsymbol{\Phi})|}$, and the corresponding strategy profile  $\{(x(v,c), q(v,c)),\forall v,c \}$ will also change. This completes the proof.
\end{IEEEproof}

\section{Proof of Proposition 6}

\begin{IEEEproof}
The idea is similar to the proof of Proposition 4. We consider the Banach Fixed Point Theorem for contraction mapping. The function $f(\delta_k)$ is defined as
$$
f(\delta_k)=
\lambda\delta_k+(1-\lambda)\wsp\cdot h(q_k)\cdot \frac{|\widetilde{\etaex_k}(\boldsymbol{\Phi})|}{|\widetilde{\etase_k}(\boldsymbol{\Phi})|}.
$$
We next show that the function $f(\delta_k)$ is a contraction mapping. Since the user partitions in the two-dimensional plane are all line segments, according to the analysis in the scenario with undifferentiated quality. Hence, the areas $\widetilde{\etaex_k}(\boldsymbol{\Phi})$ and $\widetilde{\etase_k}(\boldsymbol{\Phi})$ are both polygons, and the areas are polynomial in terms of $\delta_k$. We take the first-order derivative of $f'(\delta_k)$, i.e., $f'(\delta_k)$. By setting a large enough step size $\lambda$, we can always have $|f'(\delta_k)|\leq 1$, since the terms in $|f'(\delta_k)|$ containing $\delta_k$ can be as small as possible when increasing $\lambda$. The difference compared with the proof of Proposition 4 is that we do not have a closed-form expression for the step size $\lambda$. Yet, we can always set a large enough $\lambda$ to obtain a contraction mapping at the cost of having more iterations. Given the initial market shares, the contraction mapping implies that the iterative process will converge to a unique stable point. This completes the proof.
\end{IEEEproof}

\section{Proof of Proposition 7}

\begin{IEEEproof}
Notice that there is a one-to-one mapping between a strategy profile $\{ (x(v,c), q(v,c)),\forall v,c \}$ and an average sharing benefit $\delta_k$. The reason is that, given any strategy profile $\{ (x(v,c), q(v,c)),\forall v,c \}$, the market shares $\widetilde{\etase_k}$, $\widetilde{\etaex_k}$, and $ \widetilde{\etanl}$ will be uniquely determined. Then we can determine the new average sharing benefit $\widetilde{\delta_k} (\boldsymbol{\Phi})$ according to Eq. (44). Therefore, if the strategy profile $\{ (x^{\ast}(v,c) , q^{\ast}(v,c)),\forall v,c \}$ is a stable point, it follows that the average sharing benefit $\delta_k=\widetilde{\Phi}_k (\boldsymbol{\Phi})=\sum_{i=1}^k \frac{\wsp\cdot h(q_i)\cdot |\widetilde{\etaex_i}(\boldsymbol{\Phi})|}{\sum_{j=i}^{K}|\widetilde{\etase_j}(\boldsymbol{\Phi})|}$. Otherwise, the average sharing benefit $\delta_k$ will deviate from  $\sum_{i=1}^k \frac{\wsp\cdot h(q_i)\cdot |\widetilde{\etaex_i}(\boldsymbol{\Phi})|}{\sum_{j=i}^{K}|\widetilde{\etase_j}(\boldsymbol{\Phi})|}$, and the corresponding strategy profile $\{(x(v,c) , q(v,c)),\forall v,c \}$ will also change. This completes the proof.
\end{IEEEproof}

\end{document}